\documentclass[unnumsec,webpdf,contemporary,large]{oup-authoring-template}%

\graphicspath{{immagini/}}

\usepackage{graphicx}
\usepackage{booktabs}
\usepackage{subfig}

\theoremstyle{thmstyleone}%
\theoremstyle{thmstyletwo}%
\newtheorem{example}{Example}%
\theoremstyle{thmstylethree}%
\newtheorem{definition}{Definition}

\newcommand{\hide}[1]{}

\usepackage[mathlines, switch]{lineno}

\begin{document}

\journaltitle{xx}
\DOI{DOI here}
\copyrightyear{2026}
\pubyear{2026}
\vol{XX}
\issue{x}
\access{Published: Date added during production}
\appnotes{Paper}

\firstpage{1}


\title[Plausibility-Driven Prioritization of Candidate Biomedical Annotations]{Plausibility-Driven Prioritization\\ of Candidate Biomedical Annotations}

\author[1]{Emanuele Cavalleri\ORCID{0000-0003-1973-5712}}
\author[1]{Miad Alavinezhad\ORCID{0009-0007-3038-0875}}
\author[1]{Dario Malchiodi\ORCID{0000-0002-7574-697X}}
\author[1,$\ast$]{Marco Mesiti\ORCID{0000-0001-5701-0080}}

\address[1]{%
\orgdiv{Department of Computer Science},
\orgname{University of Milano},
\orgaddress{%
\street{Milano},
\country{Italy}}}

\corresp[$\ast$]{Corresponding author.
\href{mailto:marco.mesiti@unimi.it}{marco.mesiti@unimi.it}}

\received{Date}{0}{Year}
\revised{Date}{0}{Year}
\accepted{Date}{0}{Year}



\abstract{
The rapid growth of biomedical knowledge has made the validation of automatically generated biological annotations a major bottleneck in biomedical curation. While computational methods can rapidly produce large numbers of candidate annotations, determining which are biologically valid still requires costly expert review. Prioritizing these candidates before manual curation has therefore become a fundamental challenge. Machine learning techniques can support this process by exploiting biomedical knowledge graphs (bioKGs), which capture biological entities and their functional associations.
In this work, we propose a framework that leverages bioKGs to estimate the plausibility of candidate annotations and guide expert curation. Starting from knowledge graph embeddings, we train relation-specific binary classifiers using a community-based negative sampling strategy to obtain reliable confidence estimates. We then introduce a family of plausibility measures that combine classifier confidence, classifier reliability, and the semantic context provided by alternative relationships involving the same pair of biological entities. Unlike conventional confidence estimation, the proposed approach explicitly accounts for multiple biologically meaningful relations that may coexist between the same entities. 
Experimental results on five large bioKGs demonstrate that the proposed negative sampling strategy consistently improves classifier robustness, increasing balanced accuracy by an average of 5.8\%. Moreover, the plausibility measures outperform classifier confidence alone, enabling more effective prioritization of candidate annotations for expert review. Overall, our results show that the use of bioKGs improves the efficiency of AI-assisted biomedical curation while preserving expert control over the final annotation assessment.
} 


\otherabstract[Graphical Abstract]{\centering
\includegraphics[width=.785\linewidth]{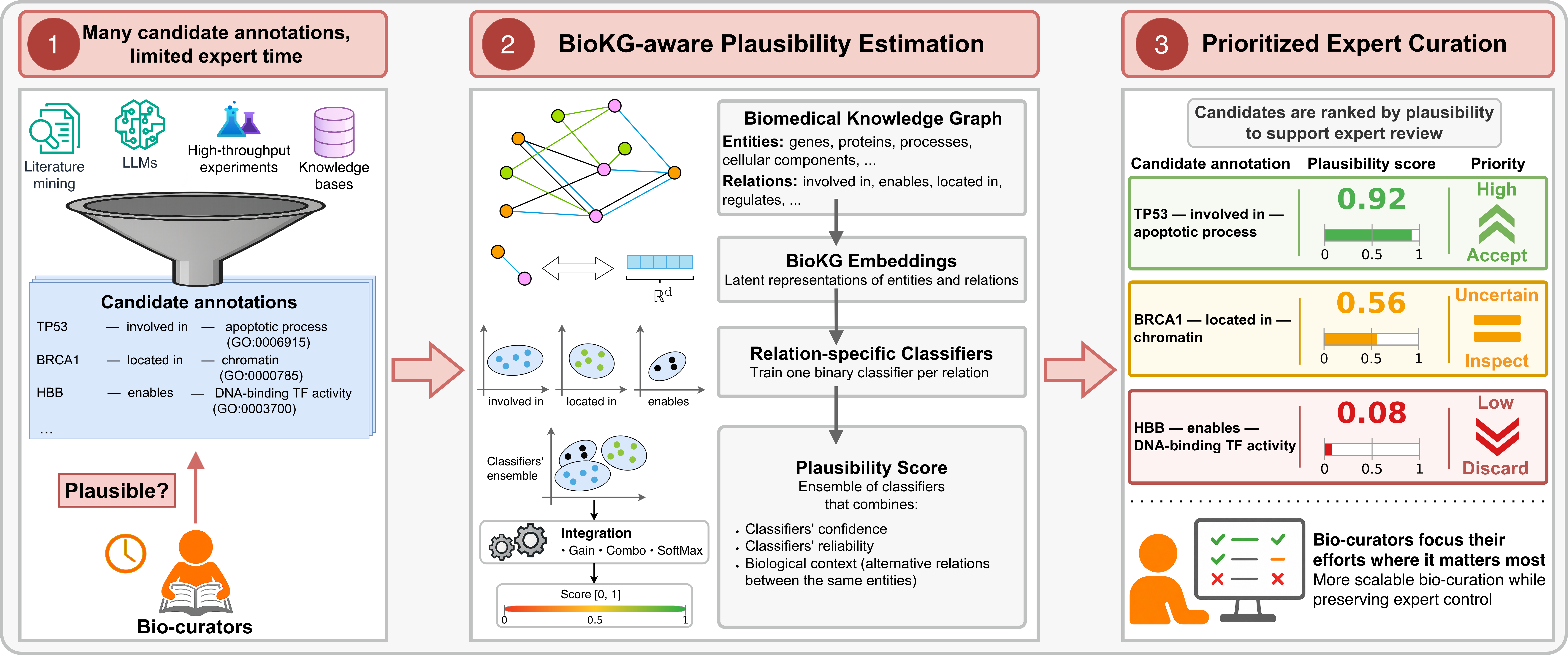}
}

\boxedtext{Key Messages}{
\begin{itemize}
\item The main challenge in modern biomedical curation is not generating candidate annotations, but efficiently supporting expert validation as biomedical knowledge continues to grow.
\item The contextual information encoded in bioKGs enables a more principled estimation of annotation plausibility than conventional confidence-based prediction.
\item AI-driven plausibility estimation can substantially improve the efficiency and scalability of biomedical curation while preserving expert control over the final validation process.
\end{itemize}}

\maketitle

\section{Introduction}
Biomedical research is generating new knowledge at an unprecedented pace, making the validation of biological annotations an increasingly critical challenge. More than 230,000 publications on non-coding RNAs have appeared since 2000, whereas only about 100--200 biocurators are responsible for transforming this continuously growing literature into structured representations~\cite{functionalAnnotation}. 
Since only a small fraction of the available knowledge can be manually curated~\cite{goflowLLM}, 
computational methods operating on Biomedical Knowledge Graphs (bioKGs)~\cite{bioKG2025} are needed to assist curators in prioritizing and validating candidate annotations while preserving expert oversight.

BioKGs provide a unified semantic representation of biomedical entities and their relationships by integrating curated databases, experimental observations, and scientific literature. Representative examples include PrimeKG~\cite{primekg}, PheKnowLator~\cite{pheno}, RNA-KG~\cite{rnakg,rnakg20}, Hetionet~\cite{hetionet}, and OptimusKG~\cite{optimuskg}. In bioKGs, annotations are represented as triples $(s,p,t)$, asserting that predicate $p$ holds between entities $s$ and $t$. For example, $(\textit{miR-21},\textit{over\mbox{-}expressed\mbox{-}in},\textit{breast cancer})$ states that microRNA \emph{miR-21} is over-expressed in \emph{breast cancer}.

Candidate annotations may be generated by literature mining systems~\cite{EuropePMC2025,PubTator}, LLMs~\cite{spires,caufield2024curategpt,goflowLLM}, high-throughput experiments~\cite{ENCODE2020}, data integration pipelines~\cite{pheno,rnakg}, or Knowledge Graph Embedding (KGE) methods~\cite{graphReprLearning}. Regardless of their origin, these annotations require expert validation before inclusion in curated bioKGs. Consequently, recent AI-assisted biocuration systems focus on supporting expert review rather than replacing human judgement~\cite{goflowLLM}.

Among the available techniques, KGEs 
are particularly 
suitable
because they learn low-dimensional representations capturing the structural and semantic information encoded in bioKGs. Originally introduced for KG completion~\cite{KGC2020}, these embeddings provide effective features for learning relation-specific classifiers that estimate the confidence of candidate annotations. However, confidence alone is insufficient because candidate annotations should also be evaluated in the context of alternative relationships involving the same entities. We therefore introduce the broader notion of \emph{plausibility}, which combines classifier confidence with contextual evidence encoded in the bioKG to provide a stronger basis for prioritizing expert review.

We propose a framework that exploits 
bioKGs to support expert curation through plausibility estimation. Starting from graph embeddings, we train relation-specific binary classifiers on homogeneous graph partitions using a community-based negative sampling strategy~\cite{glow26}. We then introduce plausibility measures that extend classifier confidence by incorporating classifier reliability and evidence provided by competing predicates between the same entity pair. The resulting scores distinguish highly plausible, implausible, and uncertain candidate annotations, enabling more effective prioritization for expert review.

Experiments on five 
large bioKGs show the effectiveness of the proposed framework. community-based negative sampling improves balanced accuracy by 5.8\% on average and yields higher confidence on unseen positive annotations. Moreover, the 
plausibility measures better discriminate target predicates from competing alternatives, improving candidate annotation prioritization
while preserving expert control over the final validation process.

Our main contributions are:

\begin{itemize}
\item a bioKG-based framework for estimating the plausibility of candidate biomedical annotations;
\item a community-based negative sampling strategy that improves the robustness of relation-specific classifiers;
\item three 
plausibility formulations combining classifier confidence, classifier reliability, and competing predicates;
\item a set of evaluation metrics for assessing plausibility formulations;
\item an extensive experimental validation on five large bioKGs.
\end{itemize}

\vspace*{-9pt}
\section{Related Work}\label{sec:RW}

KGE models~\cite{graphReprLearning} learn low-dimensional representations of entities and relations that preserve the structural and semantic information 
of
a KG. These representations have proved effective for several biomedical tasks (e.g. gene--disease association prediction~\cite{Paliwal2020}, drug interaction prediction~\cite{gemma2024}, and drug repurposing~\cite {drugreporposing}).
Early methods (e.g. DeepWalk~\cite{deepwalk}, node2vec~\cite{node2VecPaper}) exploit random walks for KGE, while more recent approaches learn embeddings for entities and relations relying on their types. Representative approaches include translation-based methods (e.g. TransE~\cite{transE}, TransH~\cite{transH}), bilinear models (e.g. DistMult~\cite{distmult}), and complex-valued formulations (e.g. ComplEx~\cite{complex}, RotatE~\cite{rotate}). 
In this paper, we use KGEs only to derive 
latent
representations that serve 
classifiers for confidence and plausibility assessment.

Binary classifiers have been widely adopted to estimate candidate annotations~\cite{KGC-survey23} through confidence scores. 
%
Several approaches formulate triple validation as a supervised classification problem. 
For example, KGBoost~\cite{KGBoost} relies on gradient-boosted decision trees, KG-BERT~\cite{KG-bert} exploits transformer representations, and recent two-stage architectures~\cite{ReDistLP,KGR3} refine link prediction outputs through supervised classification.
In this paper, we employ binary rather than multiclass classifiers because multiple biologically meaningful relations may exist between the same pair of entities. 
Binary models estimate the confidence of each predicate independently, without assuming mutual exclusivity.

Negative sampling is essential for KG learning even if explicit negative triples are rarely available. Most existing methods generate artificial negatives during embedding learning, using strategies such as corruption-based, degree-aware, adversarial, dynamic, or influence-based sampling~\cite{negativeOnEmbeddings,negativeOnEmbeddings3,negativeOnEmbeddings2,divine}.
A smaller body of work investigates negative sampling for downstream supervised classifiers~\cite{Cappelletti2024}, where the objective is to improve classification performance rather than embedding quality. 
Our work also targets downstream classification but exploits the heterogeneous semantics of bioKGs. 
By considering relation-specific communities \cite{glow26}, the proposed approach generates stronger negative annotations, improving the robustness of the confidence estimators.





\begin{figure*}[t]
    \centering
    \subfloat[Knowledge Graph $G$\label{fig:partition-kg}]{
        \includegraphics[width=0.38\textwidth]{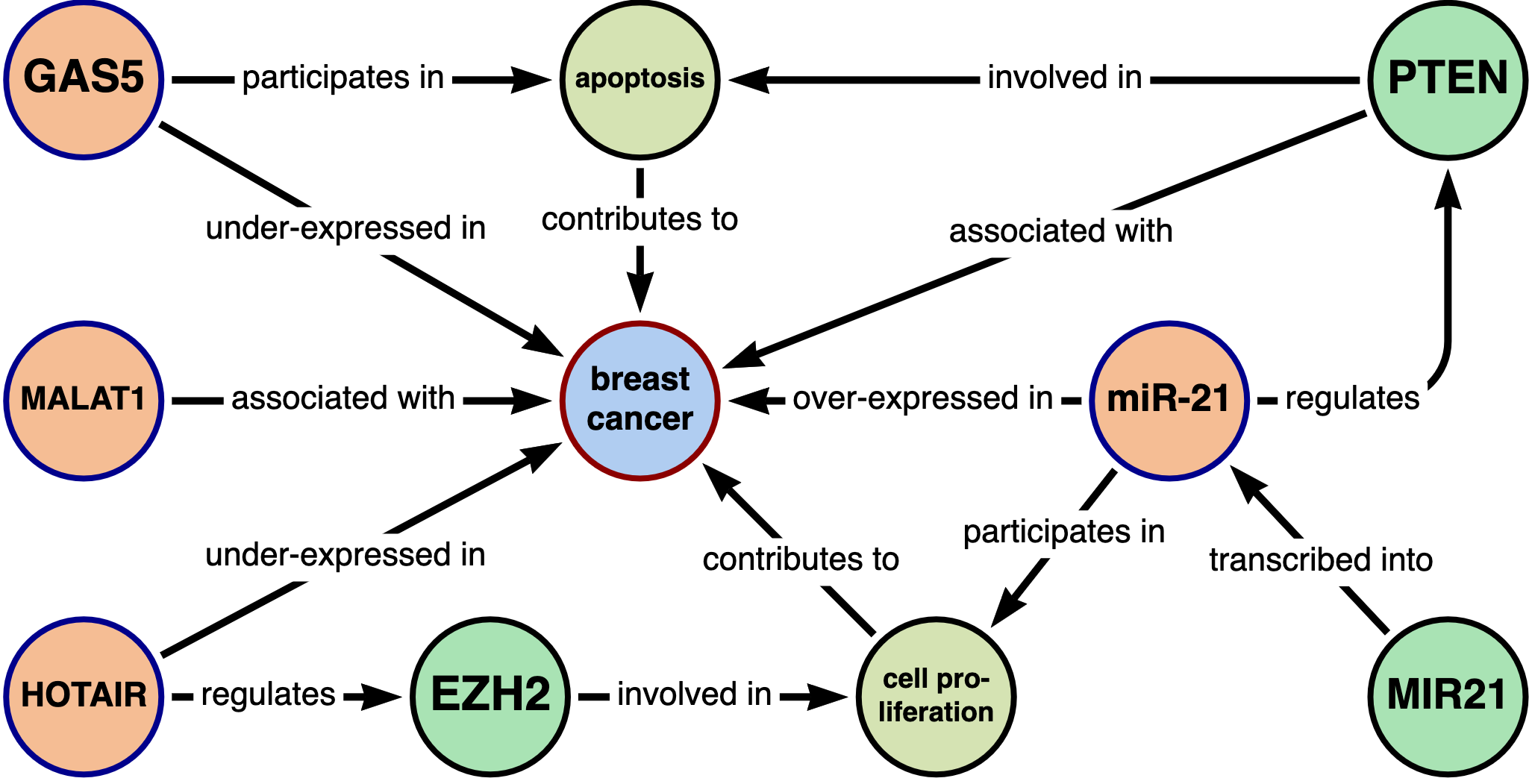}
    }
    \hfill
    \subfloat[Schema $G_S$\label{fig:partition-schema}]{
        \includegraphics[width=0.25\textwidth]{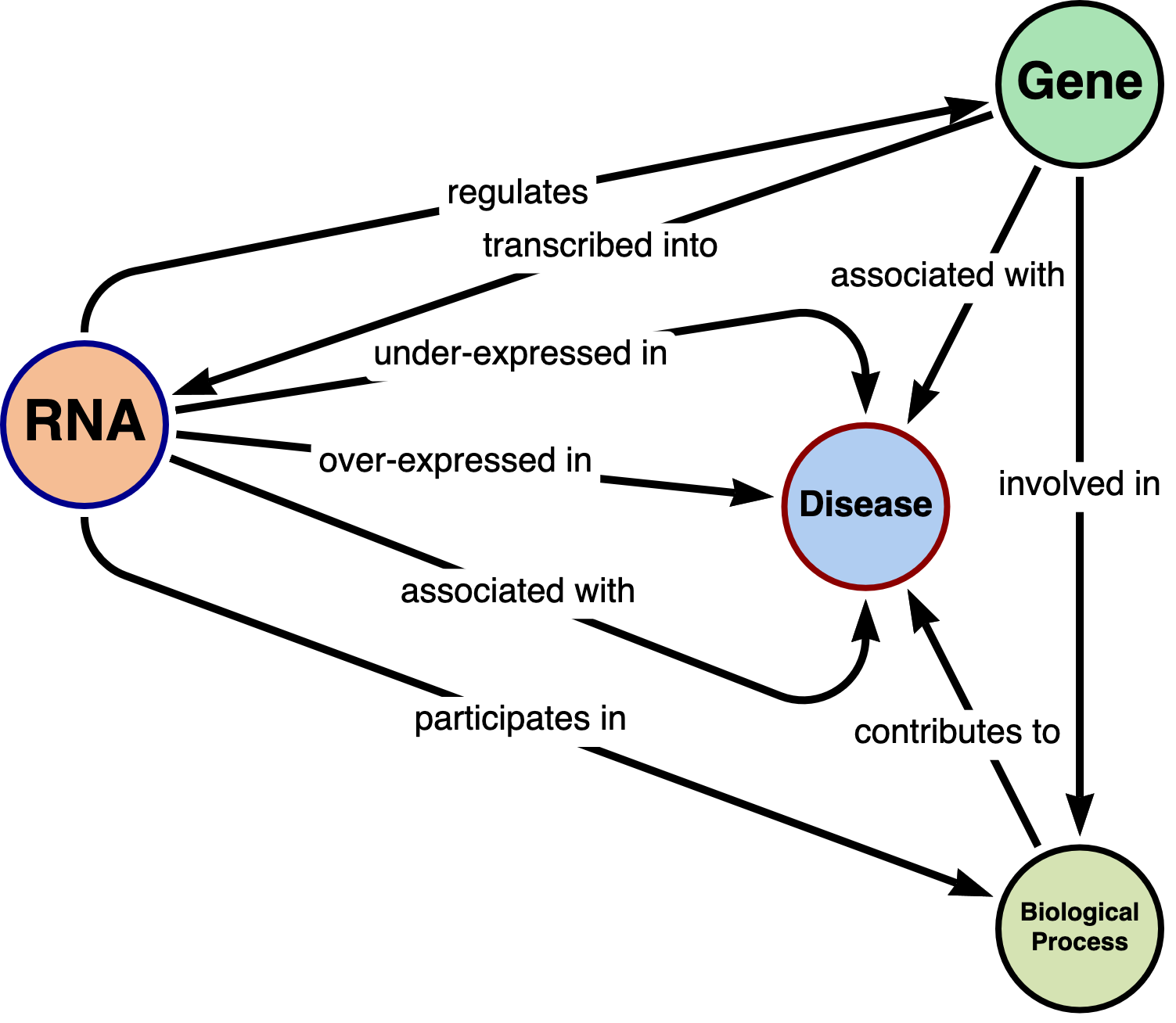}
    }
    \hfill
    \subfloat[Partition of $G$ by schema facts\label{fig:partition-partition}]{
        \includegraphics[width=0.32\textwidth]{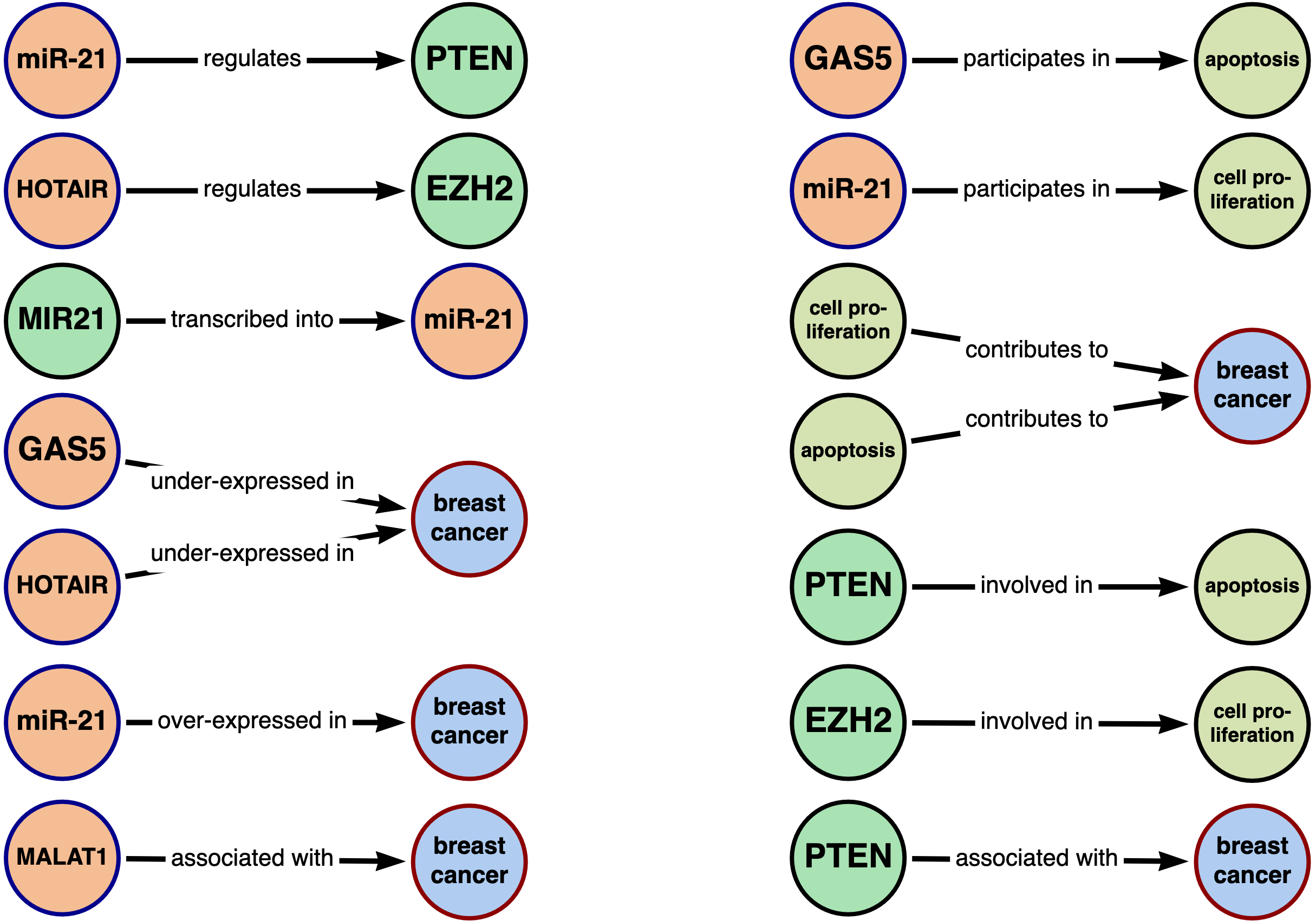}
    }

    \caption{Example of partitioning a KG $G$ according to its schema $G_S$.}
    \label{fig:partition}
\end{figure*}
\vspace*{-9pt}
\section{Preliminary Concepts}\label{sec:kg}

\subsection{Biomedical Knowledge Graphs}

Schemas provide the semantic structure of bioKGs by defining entity types (e.g. genes, drugs, diseases, 
pathways) and relation types (e.g. \emph{treats}, \emph{regulates}, \emph{associated-with}), ensuring semantic consistency.


\vspace{-3pt}

\begin{definition} ({\em KG Schema}).
Let \(N_T\) be the set of node types and \(E_T\) the set of edge types (predicates).
A {\em KG schema} is a graph
$G_S=\langle N_T,E_S,E_T\rangle,$
where
$E_S=\{(s_T,p,t_T)\mid s_T,t_T\in N_T,\ p\in E_T\}$
is the set of schema facts (denoted $\mathrm{Facts}(G_S)$).
\end{definition}

A schema acts as a blueprint for the construction of 
KGs. 

\begin{definition} ({\em KG}).
Let $G_S=\langle N_T,E_S,E_T\rangle$ 
be a KG schema. A {\em Knowledge Graph} conforming to $G_S$ is a triple $G=\langle N,E, \phi\rangle$, where:
$N$ is the set of graph nodes (entities), 
$\phi$ a function assigning a type to each node, and $E\subseteq N\times E_T\times N$ is the set of typed edges, namely
\(
E=\{(s,p,t)|\phi(s),\phi(t)\in N_T,\ p\in E_T, (\phi(s),p,\phi(t))\in E_S\}.
\)
\end{definition}

Given a KG $G$, we denote by $\mathrm{Nodes}(G)$ and $\mathrm{Edges}(G)$ its sets of nodes and edges. For a pair of entities $s,t\in N$, we define
\(
\mathrm{Facts}(s,t)=\{f\in \mathrm{Facts}(G_S)| f=(\phi(s),p,\phi(t))\},
\)
that is, the set of schema facts compatible with the types of $s$ and $t$.

For a schema fact $f=(s_T,p,t_T)\in \mathrm{Facts}(G_S)$, we define the {\em induced subgraph} $G_f$ as the subgraph of $G$ containing all nodes of type $s_T$ or $t_T$ and all edges labeled by predicate $p$. Although the nodes of $G_f$ may belong to two different types, all its edges share the same relation type. The collection of induced subgraphs associated with the schema facts forms a partition of $\mathrm{Edges}(G)$, namely
\(
\bigcup_{f\in\mathrm{Facts}(G_S)} \mathrm{Edges}(G_f)=E
\)
and
\(
\bigcap_{f\in\mathrm{Facts}(G_S)} \mathrm{Edges}(G_f)=\emptyset.
\)

\begin{example}
Figure~\ref{fig:partition-kg} shows a KG describing relationships among RNA molecules, genes, diseases, and biological processes. 
Its schema (Figure~\ref{fig:partition-schema}) allows multiple relation types between entities of the same type. Finally, Figure~\ref{fig:partition-partition} illustrates the induced subgraphs obtained by partitioning the KG according to the schema facts in $\mathrm{Facts}(G_S)$.
\end{example}

\subsection{KG Embedding and Binary Classifiers}\label{sec:classifier}
A KG $G$ is first embedded into a latent space using a KGE model, which captures both graph topology and relation types. The resulting edge embeddings are derived from the corresponding node embeddings (e.g. through concatenation or the Hadamard product).

For each schema fact \(f_i \in \mathrm{Facts}(G_S)\), we train a relation-specific binary classifier \(\mathrm{Model}_{f_i}\) on the corresponding edge embeddings. This partitioning produces homogeneous training sets while preserving the heterogeneous semantics of the original bioKG. 
Given a candidate annotation \((s_N,p,t_N)\), \(\mathrm{Model}_{f_i}\) estimates its compatibility with the structural and semantic patterns observed in the corresponding partition \(G_{f_i}\). Negative examples are generated during training to improve classifier robustness (see next section). 

\begin{figure}[t]
    \centering
    \includegraphics[width=0.48\textwidth]{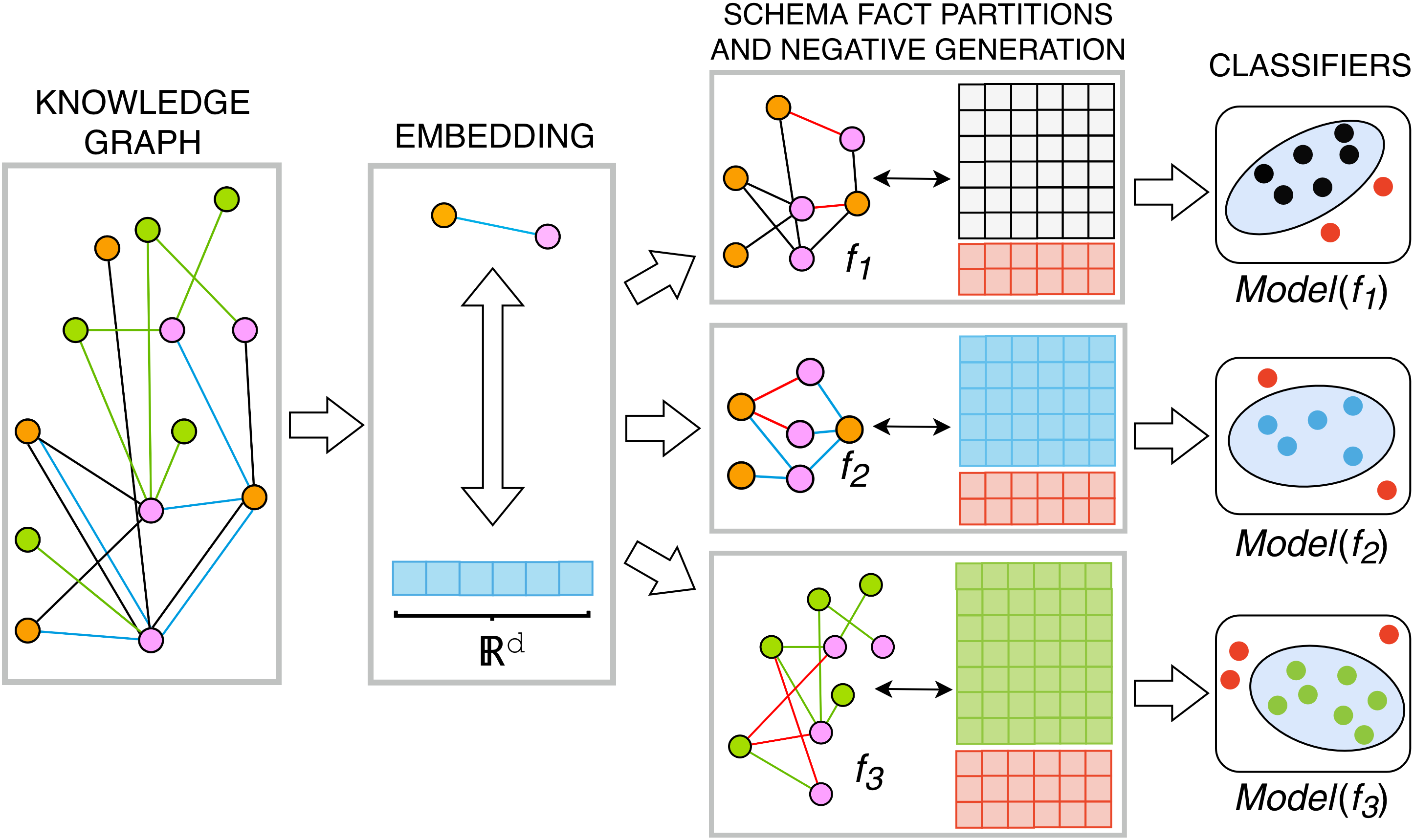}
    \caption{
    Pipeline for training classifiers in a KG.
    }
    \label{fig:pipeline_intro}
\end{figure}

Figure~\ref{fig:pipeline_intro} summarizes the training pipeline. Node and edge embeddings are first computed and then partitioned according to the schema facts. Each partition contains observed edges ($Positive(f_i)$) together with generated negative examples ($Negative(f_i)$), from which the corresponding classifier learns to distinguish plausible from implausible annotations.
Since classifier reliability depends not only on predictive performance but also on the amount of available training data, each schema fact \(f_i\) is associated with a relevance weight \(w_i\) combining the balanced accuracy \(bAcc(f_i)\) achieved on the validation set and the size of its training partition.
$w_i$ is defined as the weighted harmonic mean of the {\it normalized balanced accuracy} $ba_i$ and the {\it normalized partition size} $s_i$:
\[
w_i=
(1+\beta^2)
\frac{s_i\,ba_i}
{\beta^2 s_i+ba_i},
\]
where $\beta\!>\!1$ emphasizes balanced accuracy, whereas $\beta\!<\!1$ gives great\-er relevance to the partition size. In the formula, $ba_i$ is defined as:
\[
ba_i=
\frac{bAcc(f_i)}
{\sum_{k=1}^{n}bAcc(f_k)}.
\]
while $s_i$ is defined as:
\[
s_i=
\frac{|G_{f_i}|}
{\sum_{k=1}^{n}|G_{f_k}|}.
\]




\subsection{The Negative Selection Strategy}\label{sec:negative}

%
In~\cite{glow26},  three negative sampling strategies were compared, and the {\em community-based strategy} 
achieved the highest balanced accuracy. The method exploits the community structure of the graph induced by a schema fact $f_i$, computed through a graph clustering algorithm (e.g. Louvain~\cite{6121636}). Since nodes within the same community are more likely to be connected, negative examples are generated between nodes 
of
different communities.

\vspace{-3pt}

\begin{definition} ({\em Community-based Strategy}). Let $f_i=(s_G,p,t_G)$ be a schema fact of a KG $G$, $G_{f_i}$ the induced graph, and $\{C_1,\ldots,C_n\}$ the communities identified by a graph clustering algorithm. A negative fact $(s_N,q,t_N)\notin\mathrm{Edges}(G_{f_i})$ is added to $G_{f_i}$ iff:
$(i)$~$s_N,t_N\in\mathrm{Nodes}(G_{f_i})$, $\phi(s_N)=s_G$, $\phi(t_N)=t_G$, $q\in E_T$;
$(ii)$~$((s_N,q,t_N)\in E \land q\neq p)\lor((s_N,q,t_N)\notin E \land q=p)$;
$(iii)$~$s_N\in C_i$, $t_N\in C_j$, $i\neq j$;
$(iv)$~$|\mathrm{Negative}(G_{f_i})|\leq|\mathrm{Positive}(G_{f_i})|$.
\end{definition}

\begin{figure}[t]
    \centering
        \centering
        \includegraphics[width=0.65\linewidth]{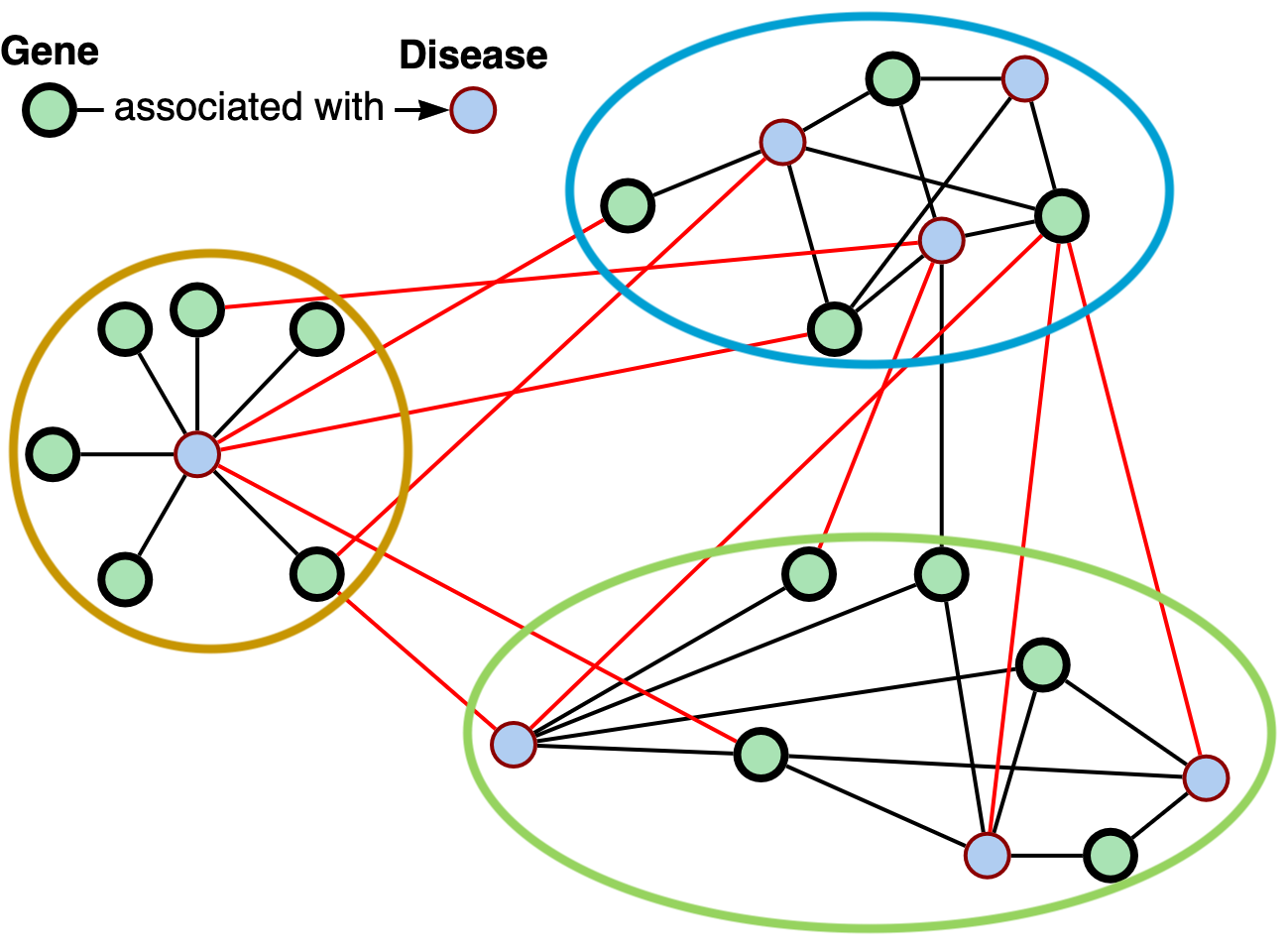}
    \caption{Community-based negative sampling strategy.}
        \label{fig:community}
\end{figure}

Figure~\ref{fig:community} illustrates the strategy on the schema fact {\em (Gene, associated-with, Disease)}. Negative examples (red) are generated between entities belonging to different communities.

\vspace{-3pt}

\section{Competition-Aware Plausibility Formulations}\label{sec:approach}

Estimating the plausibility of a candidate annotation
$(s_N,p,t_N)\notin Edges(G)$
means assessing how strongly the candidate is supported by the structural and semantic information encoded in the bioKG.
The simplest solution is to rely on the confidence assigned by the binary classifier of the corresponding schema fact. 

\begin{definition} ({\em Base Plausibility}).
Let $f_i$
be the schema fact associated with 
$(s_N, p, t_N)$, and $Model_{f_i}$ the corresponding classifier.
The {\em Base Plausibility} ${\mathcal P}_{\tt base}$ of $(s_N, p, t_N)$ is defined as:\\[-6pt]
\[{\mathcal P}_{\tt base}(s_N, p, t_N) = Model_{f_i}(s_N, p, t_N).\]
\end{definition}

Although simple and effective, this formulation 
ignores two important aspects: $(i)$ different predicates may provide alternative explanations for the same pair of entities; and $(ii)$ classifiers differ in reliability according to both their validation performance and the amount of available training data.

To address these limitations, we progressively introduce three competition-aware formulations that rely on the following base definition. Table~\ref{tab:plausibility_summary} summarizes their main characteristics. 

\begin{table}[t]
\centering
\small
\begin{tabular}{lcccl}
\toprule
Method &
Competition &
Weight &
Scope &
Purpose \\
\midrule
$\mathcal{P}_{\tt base}$ & -- & -- & -- &
Confidence estimation\\
$\mathcal{P}_{\tt gain}$ & \checkmark & \checkmark &
strongest &
Maximize predicate separation\\
$\mathcal{P}_{\tt combo}$ & \checkmark & \checkmark &
strongest &
Balance calibration and separation\\
$\mathcal{P}_{\tt soft}$ & \checkmark & \checkmark &
top-$k$ &
Multi-predicate competition\\
\bottomrule
\end{tabular}
\caption{Summary of the proposed plausibility formulations.}
\label{tab:plausibility_summary}
\end{table}
\subsection{Gain Plausibility}
Inspired by learning-to-rank approaches~\cite{liu}, we introduce a competition-aware plausibility formulation based on the concept of \emph{relative gain}, 
which measures the advantage of the target predicate over its strongest competitor. 
Since classifiers differ in reliability, each confidence score is weighted by the relevance of its schema fact, which combines classifier quality and the amount of available training data.
The relative gain of schema fact 
$f_i$ is 
defined as
\[
\begin{array}{c}
\hspace*{-1.5cm}  \Delta^{f_{i}}(s_N, p, t_N)
\!=\! 
Model_{f_i}(s_N, p, t_N)\cdot w_i     \\
\hspace*{4cm}  \!-\!
\max\limits_{j\neq i}\{Model_{f_j}(s_N,q,t_N)\cdot w_j,0\}    
\end{array}
\]
where the maximum is computed over all schema facts sharing the same source and target node types as $f_i$, but associated with a different predicate $q\neq p$. The constant $0$ ensures that the formulation remains well defined when no competing predicate exists.

Positive gain values indicate that the target predicate is preferred over its strongest competitor, whereas negative values indicate that an alternative predicate receives stronger support. Values close to zero identify cases in which competing predicates receive comparable support and therefore deserve further inspection.

\begin{definition}[Gain Plausibility]
The Gain Plausibility of the candidate annotation
$(s_N,p,t_N)$ for the schema fact
$f_i$ is expressed as

\[
\mathcal P_{\tt gain}(s_N,p,t_N)
=
\sigma\!\left(
\lambda\,
\Delta_{f_i}(s_N,p,t_N)
\right),
\]

where $\sigma(z)=\frac{1}{1+\exp(-z)}$ is the sigmoid function and
$\lambda>0$ controls the sharpness of the transformation.
\end{definition}

\subsection{Combo Plausibility}

While Gain Plausibility compares the target predicate with its strongest competitor, this comparison may become overly conservative when multiple biologically meaningful predicates receive similarly high confidence scores. In such situations, penalizing the target predicate solely because a competing predicate is also plausible may underestimate its actual biological relevance.

To alleviate this effect, we introduce \emph{Combo Plausibility}, which combines the classifier confidence with the competition-aware information provided by Gain Plausibility. The resulting formulation preserves the good calibration of the Base formulation while exploiting the relative ranking among competing predicates.

\begin{definition}[Combo Plausibility]
Let $\alpha\in[0,1]$ be a mixing parameter. The Combo Plausibility of the candidate annotation
$\bar{t}=(s_N,p,t_N)$ is defined as

\[
\mathcal{P}_{\tt combo}(\bar{t})
=
\alpha\,
\mathcal{P}_{\tt base}(\bar{t})
+
(1-\alpha)\,
\mathcal{P}_{\tt gain}(\bar{t})
\]
where $\alpha$ controls the contribution of the classifier confidence and the competition-aware component. 
\end{definition}

\begin{table*}[t]
\begin{footnotesize}
    \centering
    \begin{tabular}{|l||c|c|c|c|c|c|c|c|c|c|c|c|c|}
    \hline
    {\bf KG} & {\bf $|N|$} & {\bf $|E|$} & {\bf $|N_T|$} & {\bf most repr. $N_T$} & {\bf $|E_T|$} & {\bf most repr. schema fact} & 
    {\bf degree}$^{AVG}$
    & 
    {\bf pathLen}$^{AVG}$
    & 
    {\bf clusCoef}$^{AVG}$
    \\ \hline\hline
    {\bf miRNA-KG} 
        & $99K$ & $1.6M$ & 27 & 
        BP
        ($27K$) & 138 & 
        Disease 
        HAS 
        Phenotype
        ($374K$) & 32.62 & 5.22 & 0.11 \\ \hline
    {\bf Hetionet} 
        & $47K$ & $2.3M$ & 11 & Gene ($21K$) & 16 & 
        Genes participates BP
        ($569K$) & 95.69 & 3.60 & 0.11 \\ \hline
    {\bf PKT-KG} 
        & $560K$ & $5.5M$ & 23 & Chemical ($196K$) & 320 & 
        Disease 
        HAS
        Phenotype
        ($515K$) & 19.71 & 4.57 & 0.05 \\ \hline
    {\bf PrimeKG} 
        & $129K$ & $8.1M$ & 10 & 
        BP
        ($29K$) & 18 & 
        Drug SYN\_INT Drug
        ($2.7M$) & 125.29 & 4.86 & 0.08 \\ \hline
    {\bf OptimusKG} 
        & $191K$ & $21.8M$ & 10 & Gene ($61K$) & 36 & 
        Disease associated-with Gene
        ($9.7M$) & 228.95 & 4.01 & 0.07 \\ \hline
    \end{tabular}
    
    Abbreviations: BP (Biological Process); SYN\_INT (synergistic-interaction); HAS (has-phenotype). 
    \end{footnotesize}
    \caption{Datasets.}
    \label{tab:datasets-transposed}
\end{table*}





\subsection{SoftMax Plausibility}

To capture a broader competitive context, where 
several alternative predicates can co-exist,
we introduce \emph{SoftMax Plausibility}, which compares the target predicate with the top-$k$ competing predicates through a softmax normalization. 
\begin{definition} ({\em SoftMax Plausibility}).
The {\em SoftMax Plausibility} of \(\bar{t}=(s_N,p,t_N)\) according to \(f_i\) is:
\[
\mathcal{P}_{\tt soft}
(\bar{t})
\!=\!
\dfrac{
\exp\!\left(\lambda\!\cdot\! Model_{f_i}(s_N,p,t_N)\!\cdot\! w_i\right)
}
{
\sum\limits_{f_j \in S_k^{s_N,t_N}\cup \{f_i\}}
\exp\!\left(\lambda\!\cdot\! Model_{f_j}(s_N,p_j,t_N)\!\cdot\! w_j\right)
}
\]
where $S_k^{s_N,t_N}=\arg\max_{f_i\in Fact(s_N,t_N),|S|=k}
Model_{f_i}(s_N,p_i,t_N)$ is the set of the $top_k$ schema facts with the highest scores, and \(\lambda>0\) controls the sharpness of the softmax normalization. 
\end{definition}

\section{Experiments}
\label{sec:experiments}
We evaluated the proposed framework on five large bioKGs to assess: $(i)$ the effectiveness of the community-based negative sampling strategy; $(ii)$ the impact of classifier and embedding choices; $(iii)$ the proposed plausibility formulations; and $(iv)$ the usefulness of plausibility scores in a real 
scenario.

Node embeddings have dimension 32, edge embeddings are obtained via the Hadamard product, and RF and MLP classifiers are used. Hyperparameters were selected 
via nested cross-validation
.




\subsection{Datasets}~\label{sec:datasets}

\vspace*{-6pt}

Table~\ref{tab:datasets-transposed} summarizes the five bioKGs used in our evaluation, together with their main structural characteristics. The datasets cover complementary biomedical domains, including gene--disease associations, RNA interactions, drug--disease relationships, phenotypes, and biological processes, and range from 99K to 560K nodes and from 1.6M to 21.8M edges.
The considered bioKGs exhibit typical properties of large biological networks, including small average path lengths, low clustering coefficients, and highly skewed degree distributions (details in the Supplementary Materials).

To ensure meaningful classification tasks, we excluded schema facts containing fewer than 1000 observed triples or representing purely ontological assertions (e.g. \emph{gene,is-a,protein-coding gene}). After filtering, we retained 12 schema facts for miRNA-KG, 21 for Hetionet, 20 for PKT-KG, 22 for PrimeKG, and 23 for OptimusKG. For partitions containing more than 1.5M edges, we randomly sampled 1.5M triples to keep the computational cost manageable while preserving representative training data.

\subsection{Experimental Protocol}
\label{sec:protocol}

For each schema fact \(f_i\), the observed triples were first divided into two disjoint subsets. A random \(10\%\) was held out before training to build the blind positive test set \(BlP_i\), which is used exclusively for plausibility evaluation. Since bioKGs do not contain explicit negative facts, an equally sized set of implausible triples \(N_i\) was generated through the adopted negative sampling strategy. Neither \(BlP_i\) nor \(N_i\) 
is used during training, validation, or model selection.
The remaining \(90\%\) of observed triples were used to train the relation-specific classifiers (using 70-30\% splitting to train/validate). 
For each schema fact, negative training examples were generated either through random sampling or through the proposed community-based strategy, depending on the experiment.
This protocol separates classifier training from plausibility evaluation.

\vspace*{-3pt}

\subsection{Metrics}~\label{sec:metrics}

\vspace*{-18pt}
\subsubsection{Metrics for evaluating classifiers} 

Classifier performance is evaluated using standard metrics, including precision, recall, specificity, balanced accuracy, Matthews correlation coefficient, and the $F_\beta$-score.
In addition, we introduce the $\hat{F}_\beta$-score, which jointly penalizes false positive and false negative rates:
\(
\hat{F}_\beta
=
(1+\beta^2)
\frac{FPR \cdot FNR}
{\beta^2FNR+FPR}.
\)
Unlike the traditional $F_\beta$-score, which balances precision and recall, $\hat{F}_\beta$ directly measures the trade-off between false assertions and missed positives. Throughout the experiments, we use $\beta=1$, giving equal importance to the two error types. Hyperparameters are selected through nested cross-validation using $\hat{F}_1$ as the optimization objective. Details of the hyperparameter search are reported in the Supplementary Materials.


\subsubsection{Metrics for evaluating plausibility}
A plausibility formulation should $(i)$ distinguish unseen plausible triples from generated negatives and $(ii)$ discriminate the target predicate from alternative predicates involving the same entity pair. Accordingly, we introduce two complementary families of metrics: {\em calibration metrics} and {\em competition metrics}. 
%
%
All metrics are first computed for each schema fact $f_i$ and then averaged to obtain a global evaluation over the entire KG.

\paragraph{Calibration metrics}
Calibration measures the ability of a plausibility formulation to assign high scores to plausible triples and low scores to implausible ones.

The \emph{Positive Acceptance Rate} (PAR) measures the average plausibility assigned to the blind test set:
\[
PAR_i=
\frac{1}{|BlP_i|}
\sum_{(s,p,t)\in BlP_i}
\mathcal P(s,p,t).
\]
Conversely, the \emph{Negative Rejection Rate} (NRR) measures the average rejection of generated negatives:
\[
NRR_i=
1-
\frac1{|N_i|}
\sum_{(s,p,t)\in N_i}
\mathcal P(s,p,t).
\]
Finally, the \emph{Calibration Rate} combines the two quantities:
\[
Cal_i=
\frac{PAR_i+NRR_i}{2}.
\]
High values of these metrics indicate that plausible triples receive high plausibility scores while negative ones are correctly rejected.

\paragraph{Competition metrics}
Calibration alone is insufficient when multiple biologically meaningful predicates may connect the same pair of entities. A plausibility formulation should also distinguish the target predicate from competing alternatives. To assess this property, we introduce three competition-oriented metrics.

    \begin{figure}[t]
    \centering
    \includegraphics[width=0.85\linewidth]{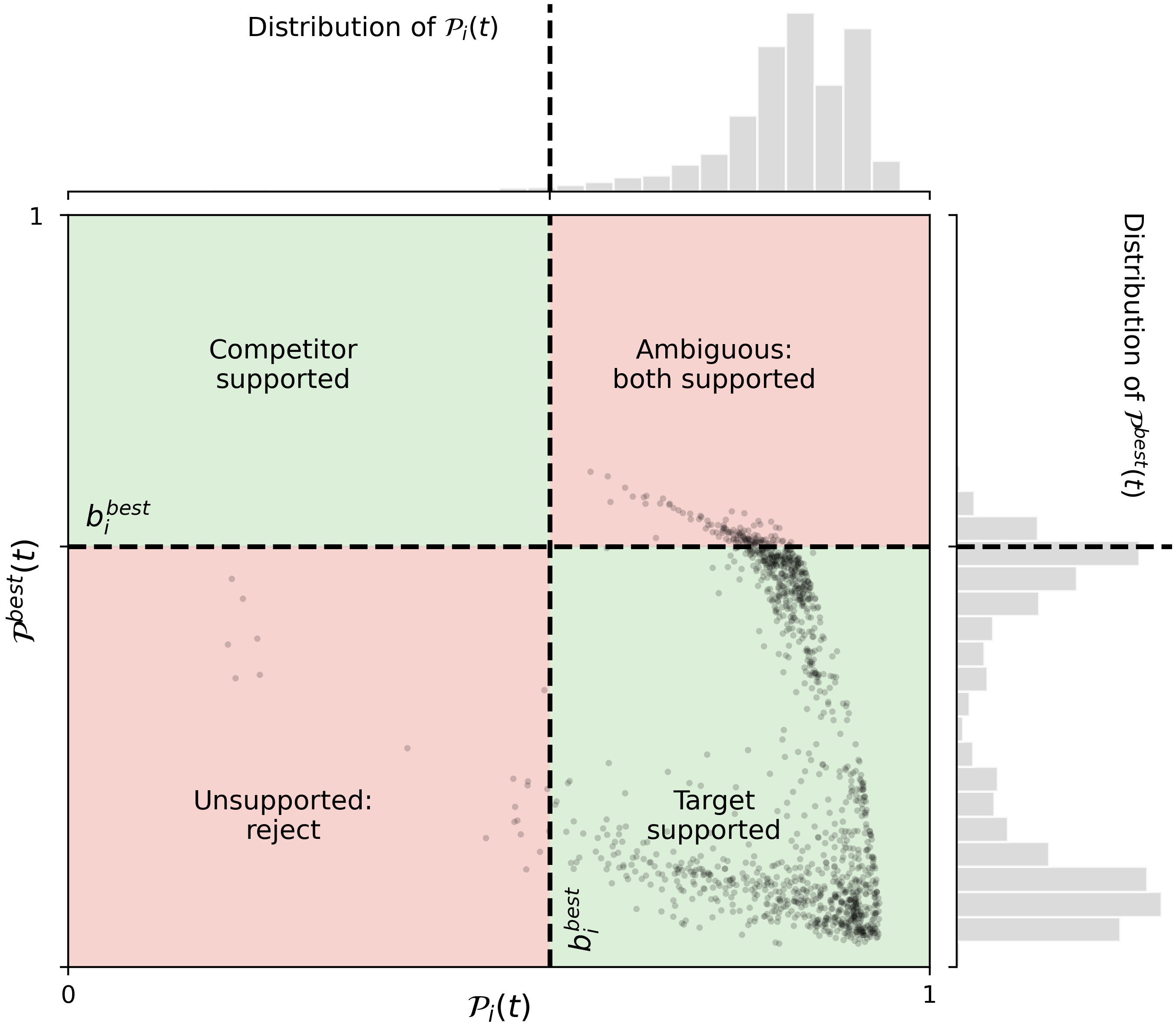}
    \caption{
    Graphical interpretation of \textit{best-competitor separation rate}. Green regions indicate triples for which the target and competitor scores fall on opposite sides of the threshold, contributing \(1\) to the metric; red regions indicate triples for which both scores fall on the same side, contributing \(0\).
    }
    \label{fig:best_competitor_metric}
\end{figure}

The \emph{Average Competitor Separation} measures the average plausibility difference between the target predicate and all competing predicates associated with the same entity pair:
\[
Sep_i =
\frac{1}{|BlP_i||C_i|}
\sum_{(s,p,t)\in BlP_i}
\sum_{p_j\in C_i}
\left|
\mathcal P(s,p,t)
-
\mathcal P(s,p_j,t)
\right|.
\]
Higher values indicate better separation of competing predicates.

The previous metric equally weights all competing predicates. However, in practice the strongest competitor often represents the most informative alternative explanation for a candidate triple.
Let
\[
\mathcal P^{best}(s,p,t)
=
\max_{p_j\in C_i}
\mathcal P(s,p_j,t)
\]
denote the plausibility assigned by the strongest competing predicate. We define the schema-fact-specific threshold
\[
b_i^{best}
=
\frac{
PAR_i
+
PAR_i^{best}
}{2}.
\]
The \emph{Best-Competitor Separation Rate} is
\[
Sep_i^{best}
=
\frac1{|BlP_i|}
\sum_{\bar t\in BlP_i}
\mathbb I
\left[
(\mathcal P(\bar t)\ge b_i^{best})
\oplus
(\mathcal P^{best}(\bar t)\ge b_i^{best})
\right].
\]
It measures the fraction of triples for which the target predicate and its strongest competitor fall on opposite sides of the threshold.

To evaluate whether the target predicate remains distinguishable even from the least plausible competing alternative, $P^{worst}$ is introduced to identify the minimal plausibility 
($b_i^{worst}$ and $Sep_i^{worst}$ are defined accordingly). 



Figure~\ref{fig:best_competitor_metric} shows the interpretation of $Sep^{best}$. Each point represents a blind plausible triple according to the plausibility assigned by the target predicate and its strongest competitor. The threshold partitions the space into four regions. Triples lying in the \emph{Target/Competitor preferred} 
regions contribute positively to the metric, whereas triples in the \emph{Rejected/Ambiguous} regions indicate insufficient separation between competing predicates.

\begin{table}[t]
    \centering
    \small
    \begin{tabular}{|l|c|c|c|}

\hline
{\bf KG} & {\bf Community} & {\bf Random} & {\bf Shared}\\\hline\hline

{\bf miRNA-KG} & $147.3K$ & $255.7K$ & $5.0K$ (1.2\%)\\\hline
{\bf Hetionet} & $1.6M$ & $2.2M$ & $56.7K$ (1.5\%)\\\hline
{\bf PKT-KG} & $1.1M$ & $1.7M$ & $35.7K$ (1.3\%)\\\hline
{\bf PrimeKG} & $2.3M$ & $4.4M$ & $170.8K$ (2.6\%)\\\hline
{\bf OptimusKG} & $3.7M$ & $6.1M$ & $336.6K$ (3.5\%)\\\hline\hline
{\bf \tt Total} & $8.8M$ & $14.8M$ & $604.8K$ (2.6\%) \\\hline

    \end{tabular}
    \caption{Overlaps of negative edges between the community-based strategy and random baseline.
    }
    \label{tab:negativeOverlaps}
\end{table}

\subsection{Selecting the Experimental Configuration}

Before evaluating plausibility, we compare the proposed community-based negative sampling strategy with a random baseline and identify the experimental configuration adopted in the remainder of the paper.

Table~\ref{tab:negativeOverlaps} reports the overlap between the negative triples generated by the two strategies. Across the five bioKGs, at most 3.5\% of the generated negatives are shared, indicating that the community-based strategy produces substantially different training examples rather than a small variation of random sampling. Consequently, any performance differences observed in the following experiments can be attributed to the quality of the generated negatives rather than to trivial overlaps. Moreover, negative generation is computationally inexpensive, requiring less than one minute per KG partition. 

\begin{figure*}[t]
    \centering

\subfloat[Average balanced accuracy with standard deviation\label{fig:balancedAccuracies}]{
        \includegraphics[width=.31\linewidth]{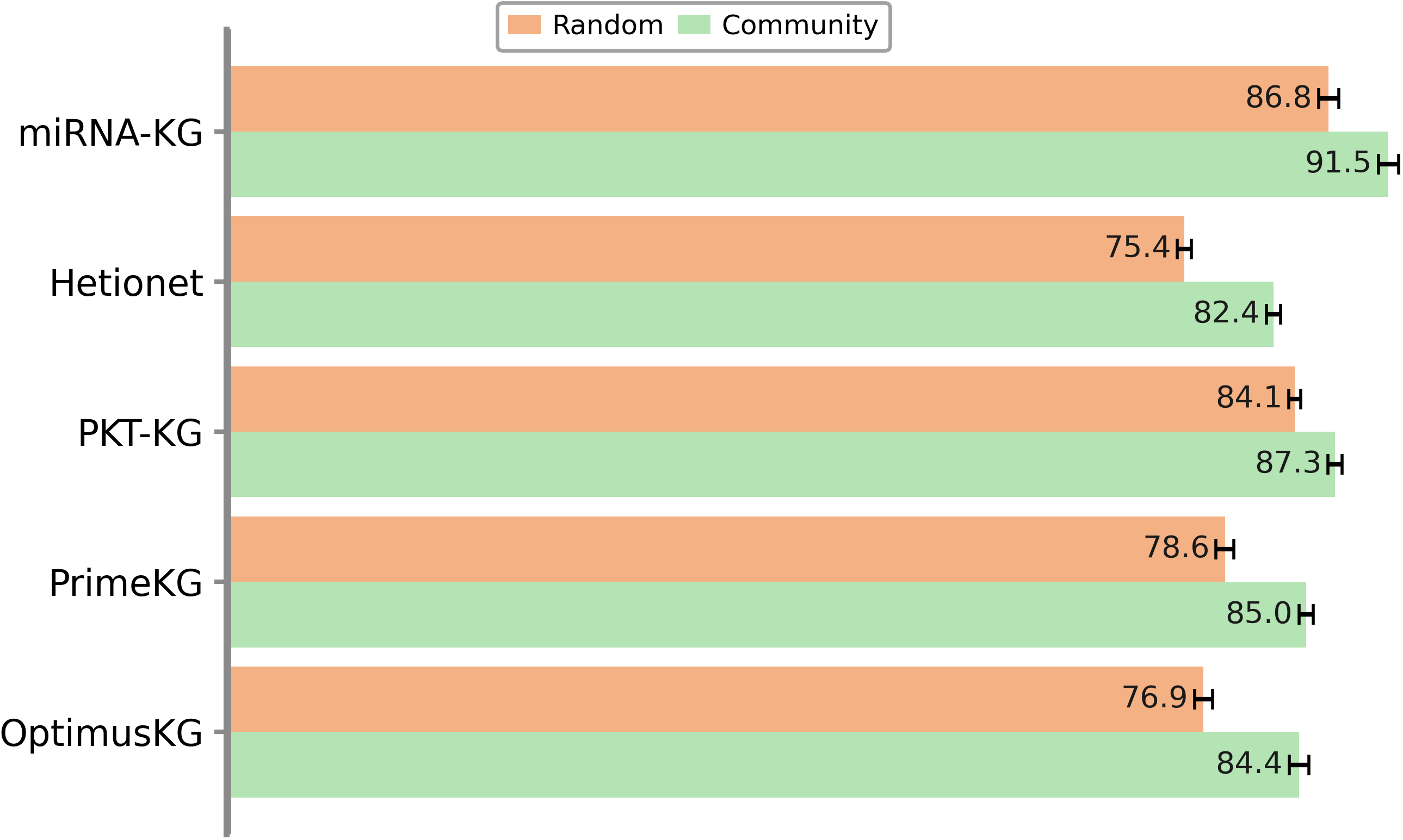}
    }
    \ \hspace{.1cm}\ 
    \subfloat[Balanced accuracy\label{fig:regression_bacc}]{
        \includegraphics[width=.31\linewidth]{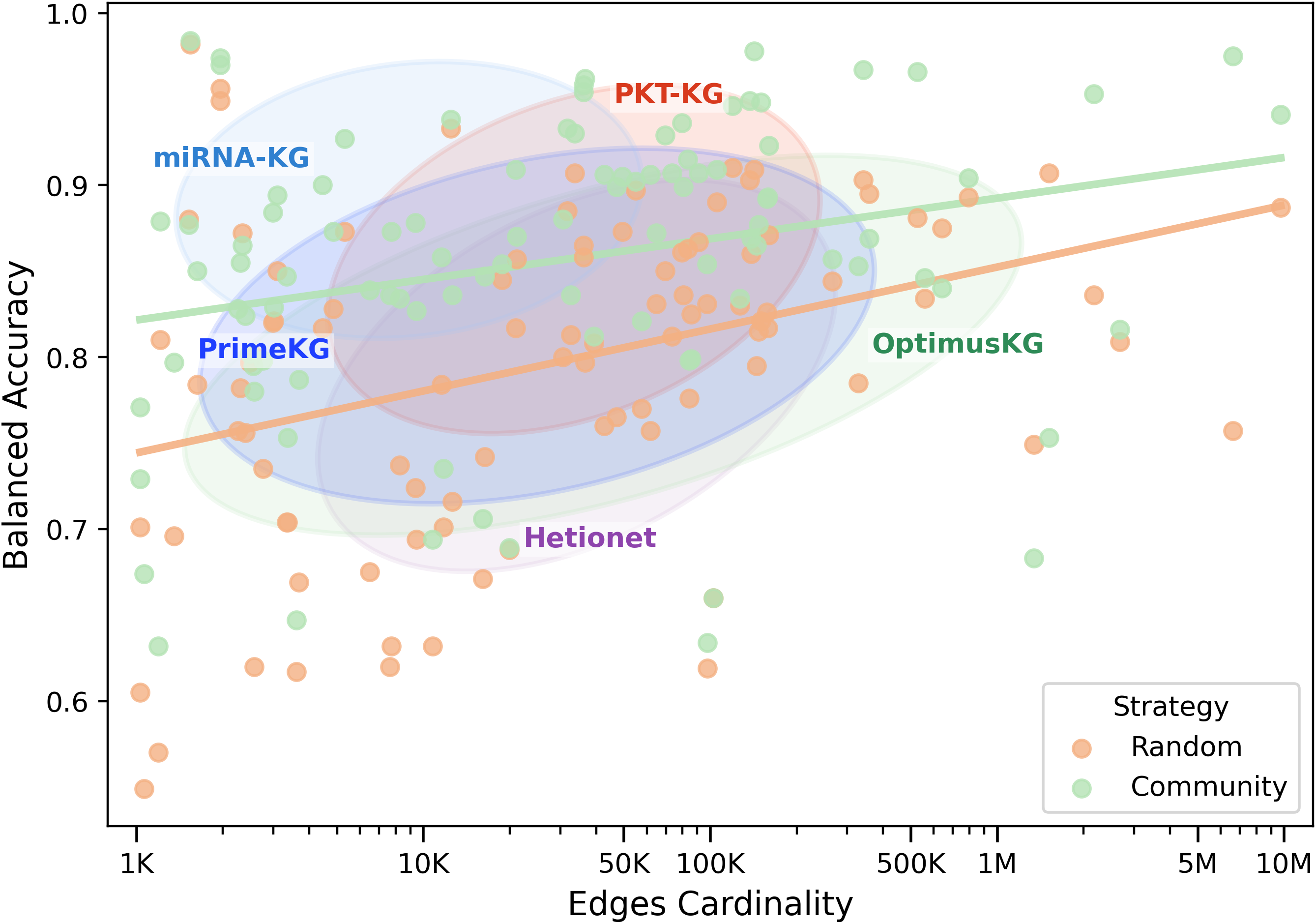}
    }
    \ \hspace{.1cm}\ 
    \subfloat[Execution time\label{fig:regression_time}]{
        \includegraphics[width=.31\linewidth]{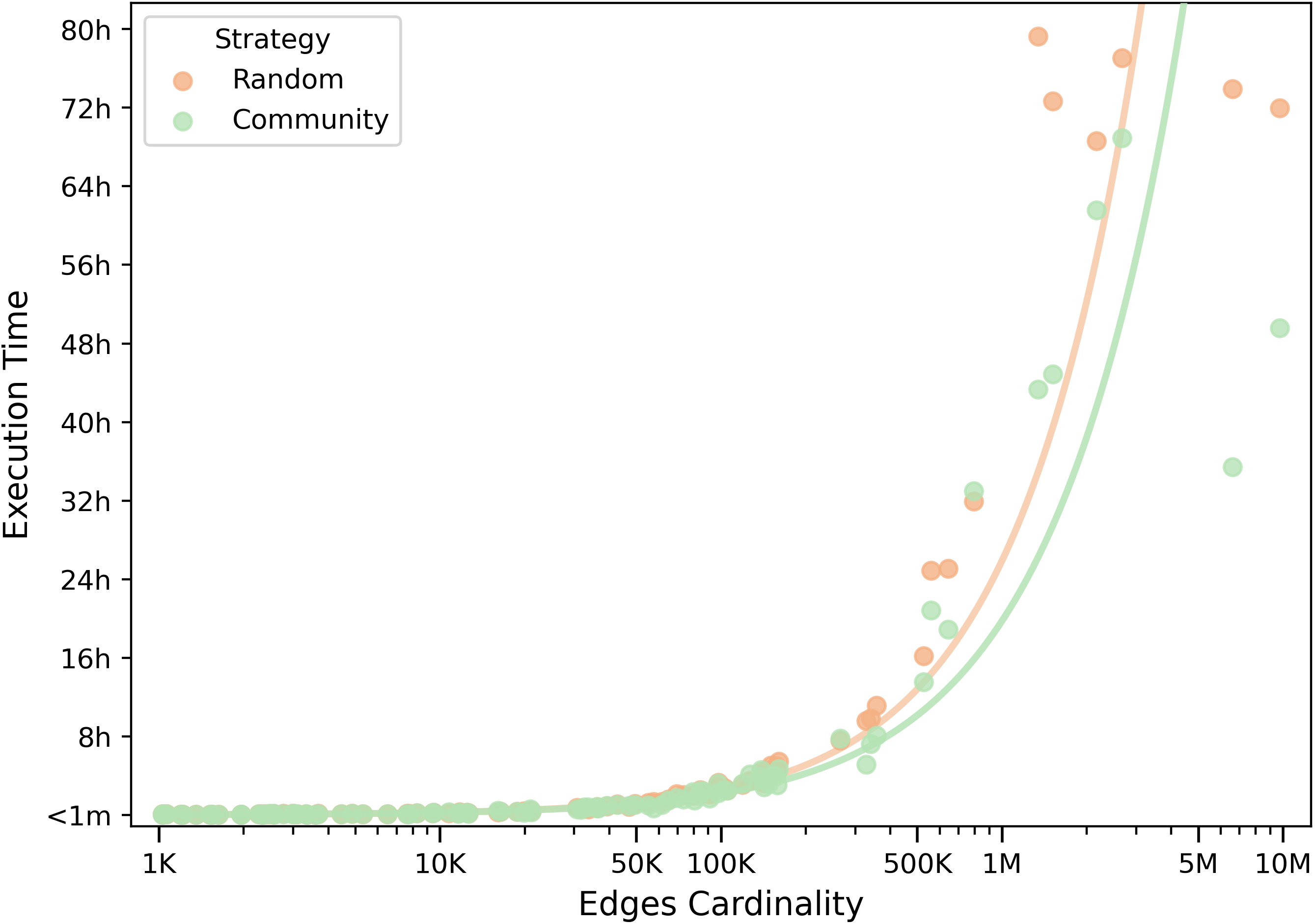}
    }

    \caption{Different perspectives of 
    classification performance. (a) Average balanced accuracy with standard deviation; (b) Regression analysis on balanced accuracy. (c) Execution time as functions of edge cardinality (log-scale) for the considered negative sampling strategies.}
    \label{fig:regression}
\end{figure*}

To determine the training configuration, we evaluated different negative-to-positive ratios and two classification models (RF and MLP) on miRNA-KG and PKT-KG. As shown in Table~\ref{tab:negtopos}, both classifiers achieve their highest balanced accuracy using a 1:1 ratio between positive and negative examples. RF consistently provides slightly higher accuracy and lower variance than MLP and is therefore adopted in the remaining experiments.

Since the proposed framework is embedding-agnostic, we also compared five heterogeneous KGE models. RotatE achieves the highest balanced accuracy, followed closely by DistMult and TransE (Table~\ref{tab:embeddingComparison}). However, TransE requires only five minutes to compute the embeddings, whereas the remaining models require several 
hours
to provide 
only marginal accuracy improvements. We therefore adopt TransE as the default embedding model in the remainder of the paper.

\begin{table}[b]
\small
\centering

\subfloat[Across negative-to-positive ratios.\label{tab:negtopos}]{
\begin{tabular}{c|ccccc}
\toprule
& \multicolumn{5}{c}{{\bf Neg-to-Pos ratio}} \\
{\bf Model} & 0.25 & 0.5 & 1 & 2 & 3 \\
\midrule
MLP & 83.8 $\pm$ 2.0 & 87.6 $\pm$ 1.4 & {\bf 88.9 $\pm$ 1.1} & 88.1 $\pm$ 1.2 & 86.6 $\pm$ 1.0 \\
RF & 84.4 $\pm$ 1.1 & 88.0 $\pm$ 0.8 & {\bf 89.4 $\pm$ 0.7} & 88.0 $\pm$ 0.6 & 84.8 $\pm$ 0.7 \\
\bottomrule
\end{tabular}
}

\vspace{0.5em}

\subfloat[Across embedding models using community-based negatives.\label{tab:embeddingComparison}]{
\begin{tabular}{l|ccccc}
\toprule
& TransE & DistMult & RotatE & ComplEx & TransH \\
\midrule
B. Acc. & 89.4 $\pm$ 0.7 & 89.5 $\pm$ 0.9 & \textbf{90.7 $\pm$ 0.7} & 81.2 $\pm$ 1.2 & 87.3 $\pm$ 0.9 \\
Time & \textbf{5} min. & 969 min. & 976 min. & 1075 min. & 1592 min. \\
\bottomrule
\end{tabular}
}

\caption{
Balanced accuracy (mean $\pm$ std) and embedding time.
Best values are reported in bold.
}
\end{table}

\subsection{Evaluating the Classification Approach}


Figure~\ref{fig:balancedAccuracies} reports the average balanced accuracy achieved by the RF classifiers across the 
five bioKGs using
two negative sampling strategies.
The community-based strategy consistently outperforms the random baseline, yielding an average improvement of \(5.8\%\), with gains observed on all datasets. The largest improvement is obtained on OptimusKG (\(+7.5\%\)), confirming 
the superiority of the community-based strategy.




Figure~\ref{fig:regression_bacc} further analyzes the relationship between classifier performance and the size of the schema-fact partitions. Balanced accuracy generally increases with the number of positive triples available for training, indicating that larger partitions provide richer structural information for learning robust classifiers. Across the entire range of partition sizes, the community-based strategy consistently achieves higher performance than random sampling. Detailed results for each schema fact are reported in the Supplementary Materials.

To assess whether the observed improvements are statistically significant, we applied both a chi-square test of independence and McNemar's test. The analyses confirm that the community-based strategy consistently produces different and significantly more accurate predictions on unseen plausible triples. In particular, community-based classifiers assign higher confidence scores than the random baseline in \(92\) out of \(98\) schema facts (\(93.9\%\)), with an average increase of \(10.1\) percentage points. Complete statistical results are provided in the Supplementary Materials.

Finally, Figure~\ref{fig:regression_time} evaluates scalability by relating training time to partition size. The proposed strategy exhibits execution times comparable to the random baseline while maintaining a more favorable trend on the largest partitions, showing that the improved predictive performance is achieved without compromising scalability.

\subsection{Quantitative Evaluation of Plausibility Scores}

Table~\ref{tab:plausibility_eval} compares the four proposed plausibility formulations using the global calibration and competition metrics. 
Unless otherwise stated, the parameter configuration providing the best validation performance is adopted; a sensitivity analysis is reported in the Supplementary Materials.


\begin{table}[b]
\centering
\small
\begin{tabular}{lcccc}
\toprule
Metric & Base & Gain & Combo & SoftMax \\
\midrule
PAR & \textbf{80.4\% (+10.2)} & 73.2\% (-1.0) & 76.9\% (+4.6) & 63.0\% (+3.7) \\
NRR & \textbf{86.5\% (+2.5)} & 52.8\% (+5.4) & 69.7\% (+3.9) & 82.8\% (+3.9) \\
Cal & \textbf{83.4\% (+6.4)} & 63.0\% (+2.2) & 73.3\% (+4.3) & 72.9\% (+3.8) \\
Sep & 30.2\% (+0.1) & \textbf{49.7\% (+3.6)} & 37.2\% (+2.2) & 48.0\% (+3.8) \\
Sep$^{best}$ & 51.2\% (+1.1) & \textbf{99.3\% ($\pm$0.0)} & 84.9\% (+6.7) & 94.3\% (+1.2) \\
Sep$^{worst}$ & 62.5\% (-0.9) & 69.4\% (+0.8) & \textbf{77.7\% (+2.6)} & 72.3\% (+0.8) \\
\bottomrule
\end{tabular}
\caption{
Global
plausibility evaluation across all schema facts. 
Best value for each metric is highlighted in bold.
}
\label{tab:plausibility_eval}
\end{table}

For
calibration, \(\mathcal P_{\tt base}\) achieves the highest values for \(PAR\), \(NRR\), and \(Cal\), indicating that the classifier confidence alone provides the best discrimination between plausible and generated negative triples.

\begin{figure*}[t]
    \centering

    \subfloat[Community-based negatives\label{fig:bio_competitors_associates_comm}]{
        \includegraphics[width=.4\linewidth]{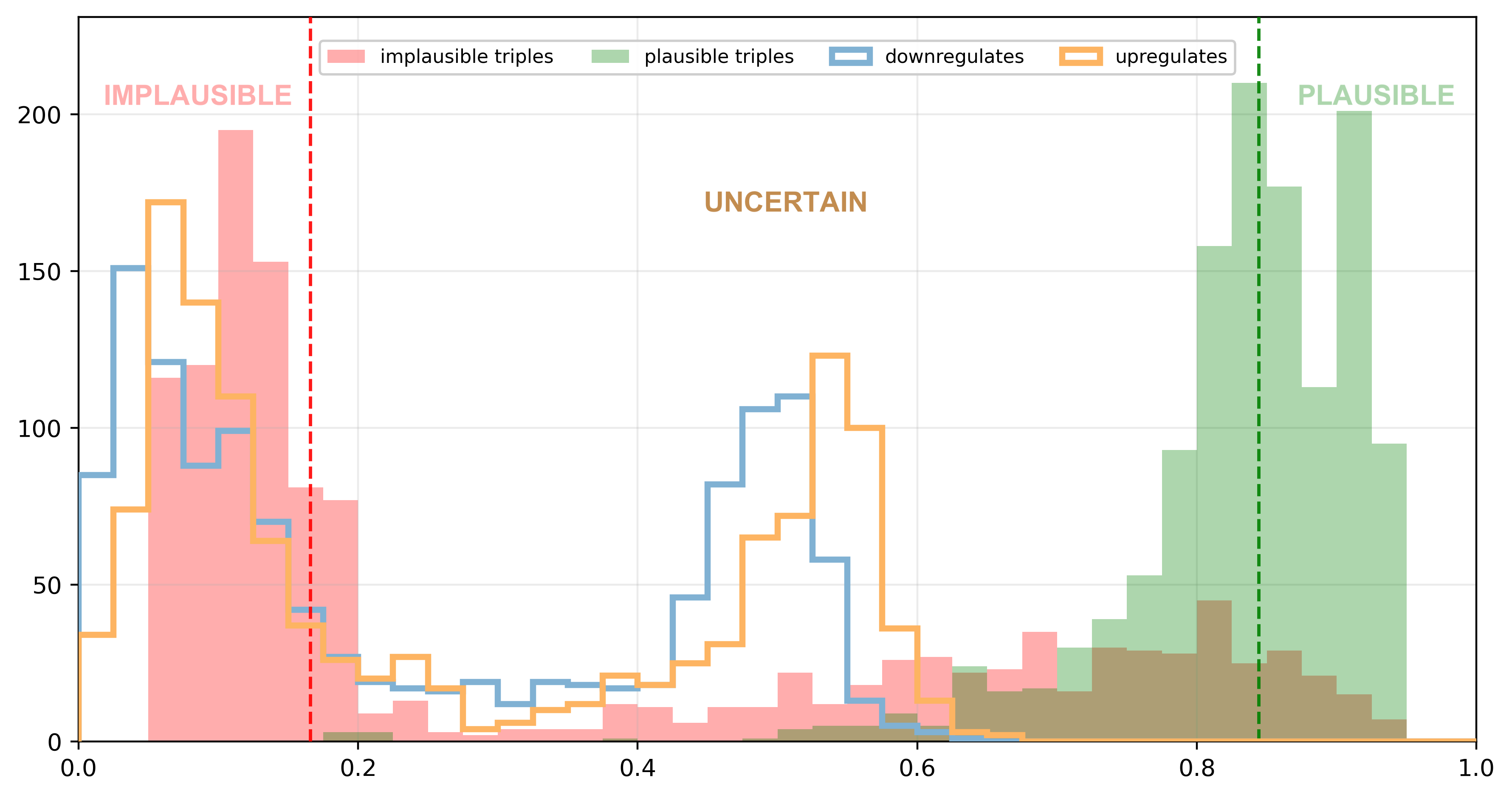}
    }
    \ \hspace{.5cm} \
    \subfloat[Random negatives\label{fig:bio_competitors_associates_rand}]{
        \includegraphics[width=.4\linewidth]{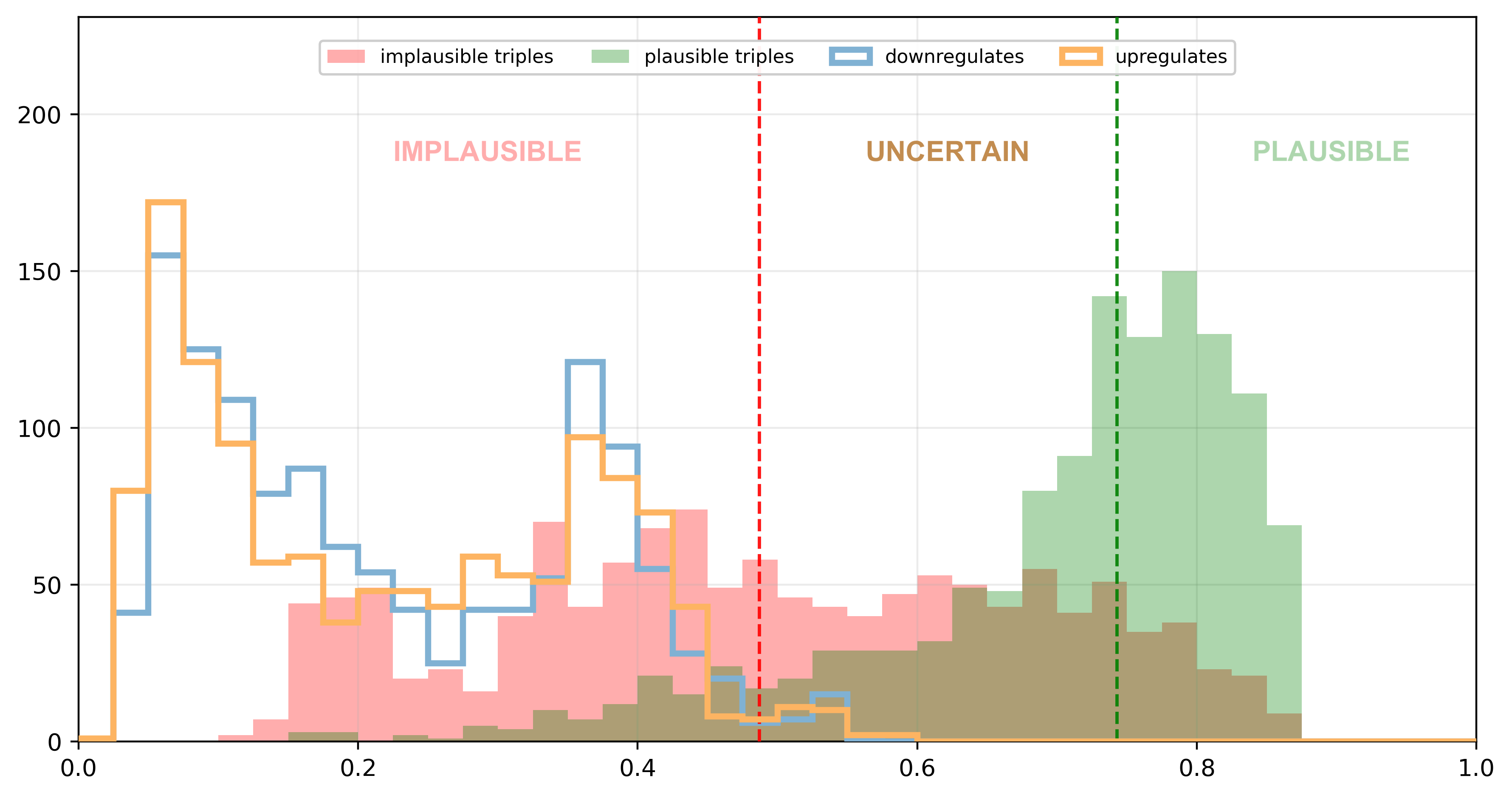}
    }

    \caption{
    Distribution of \(\mathcal{P}_{\tt combo}\) scores for the schema fact \textit{(Disease,associates,Gene)} in Hetionet. The vertical dashed lines delimit the plausibility regions.
    }
    \label{fig:bio_competitors_associates}
\end{figure*}

The competition-aware formulations 
really
improve predicate discrimination. Specifically, \(\mathcal P_{\tt gain}\) achieves the highest values for \(Sep\) and \(Sep^{best}\), confirming that comparing the target predicate with competing alternatives greatly enhances predicate-level separation.

Among the proposed formulations, \(\mathcal P_{\tt combo}\) provides the best overall compromise between calibration and competition-aware discrimination. Although it does not maximize every metric individually, it consistently achieves balanced performance across both evaluation dimensions, making it the most suitable formulation for prioritizing candidate annotations.

Overall, the quantitative evaluation shows that competition-aware plausibility formulations provide substantially richer information than classifier confidence alone. While the base formulation remains the best calibrated, the proposed competition-aware formulations markedly improve predicate discrimination, with \(\mathcal P_{\tt combo}\) providing the most balanced behaviour across the two evaluation dimensions.

\subsection{Qualitative Analysis of Plausibility Distributions}

To better understand the behaviour 
of
the proposed plausibility formulations
in realistic curation scenarios, we now inspect the plausibility distributions produced for a representative schema fact. The goal is to assess whether the proposed formulation distinguishes implausible triples, highly plausible annotations, and biologically meaningful competing predicates requiring expert inspection.

Figure~\ref{fig:bio_competitors_associates} illustrates the \(\mathcal P_{\tt combo}\) distributions obtained for the schema fact \textit{(Disease,associates,Gene)} in Hetionet. Green bars correspond to blind plausible triples, red bars to generated negatives, while the blue and orange curves represent the plausibility assigned to the same disease--gene pairs by the competing predicates \textit{downregulates} and \textit{upregulates}. Panels (a) and (b) compare classifiers trained using community-based and random negatives.

Three observations emerge.
First, community-based negatives produce a clearer separation between plausible and implausible triples. The overlap between the corresponding score distributions decreases from 49.8\% to 26.6\%, indicating that the classifier assigns more decisive plausibility estimates.
Second, competing predicates are not simply rejected. Most competitor scores concentrate in the intermediate plausibility region, reflecting that alternative disease--gene relationships may remain biologically meaningful even when they do not correspond to the target predicate.
Finally, the resulting score distribution naturally supports the curation workflow. Highly plausible triples can be prioritized for validation, implausible ones can be safely discarded, whereas intermediate scores identify candidate annotations deserving further expert inspection.

Similar behaviour was consistently observed across the remaining schema facts and bioKGs. Community-based negatives systematically reduced the overlap between plausible and implausible triples while preserving meaningful plausibility scores for biologically related competing predicates. Additional qualitative examples are reported in the Supplementary Materials.

\subsection{Case Study: Plausibility-Guided Biomedical Annotation}
\label{sec:case_study}

\begin{figure}[t]
    \centering\includegraphics[width=0.85\linewidth]{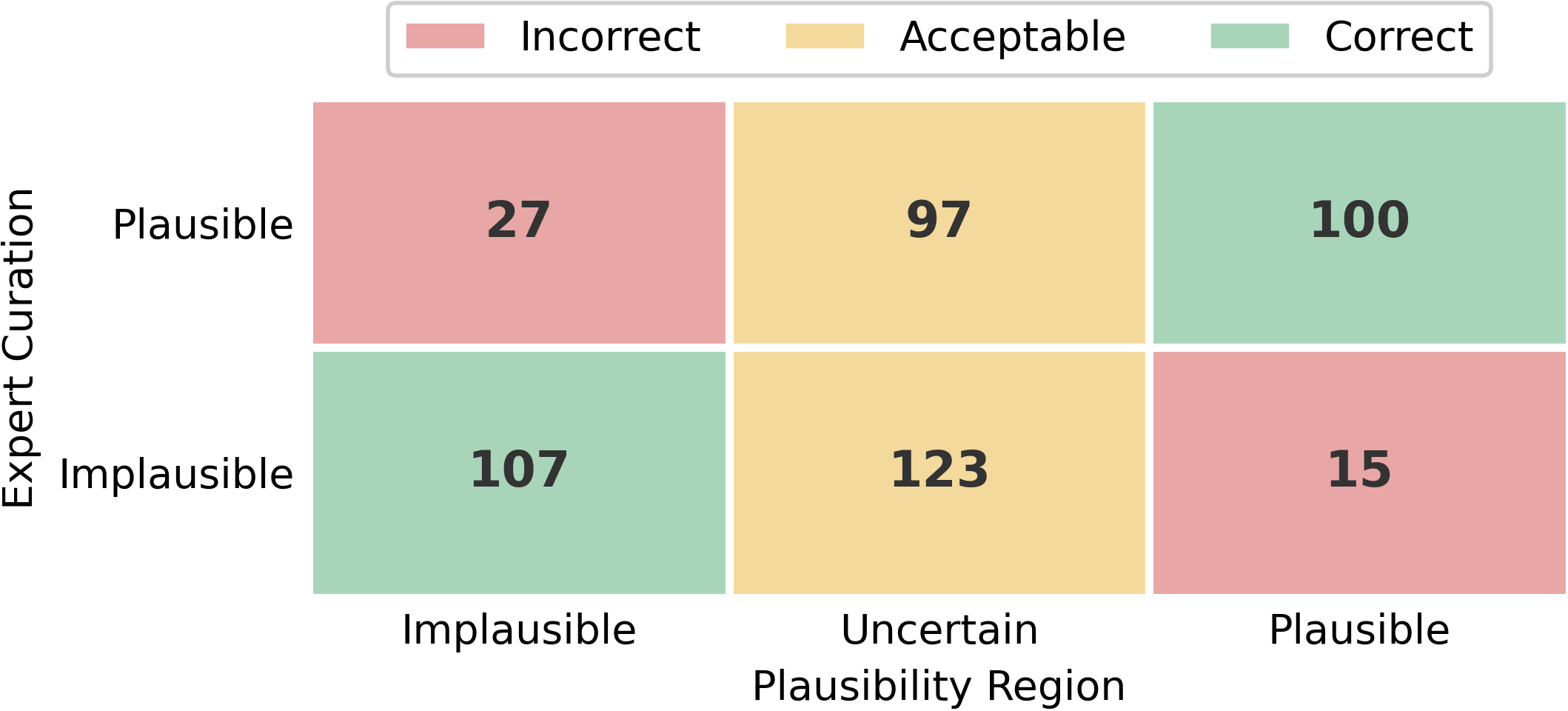}
        \caption{
        Agreement between plausibility regions and expert curation.
        }
\label{fig:case_study_plausibility_distribution}
\end{figure}

To illustrate the practical usefulness of the proposed framework, we consider a biomedical curation scenario in which candidate annotations are automatically extracted from scientific articles and subsequently reviewed by a domain expert. The objective is to evaluate whether plausibility scores can effectively prioritize candidate annotations before manual validation.

The study considers 50 scientific articles describing functional and pathway annotations involving RNAs, genes, proteins, chemicals, and environmental exposures. Candidate triples were extracted using an LLM-based relation extraction framework~\cite{cikm06}. 
Extracted entities and relations were normalized according to standard ontologies (e.g. HGNC, KEGG, MeSH, and OBO), enabling direct integration with the considered bioKGs.
Overall, the dataset contains 325 entities, 27 schema facts, and 469 candidate annotations. Each annotation was evaluated using 
\(\mathcal{P}_{\tt combo}\) and 
assessed by a biomedical expert. 

Figure~\ref{fig:case_study_plausibility_distribution} summarizes the agreement between plausibility regions and expert assessment. Overall, 207 annotations (44.1\%) fall into the agreement regions, where both plausibility and expert judgement identify the annotation as either plausible or implausible. Moreover, 220 annotations are assigned to the uncertain region. Rather than representing incorrect predictions, these cases identify candidate annotations requiring manual inspection, which is precisely the 
role of the proposed framework in supporting biomedical curation.

Only 42 annotations (9.0\%) disagree with the expert assessment.  
The disagreements are concentrated in a limited number of schema facts. In particular, \(13\) involve \textit{(miRNA,part-of,GO term)}, suggesting a mismatch between the broad functional association recognized by the curator and the stricter semantics of the \textit{part-of} predicate encoded in miRNA-KG. 
A further \(16\) disagreements involve the predicates \textit{regulates}, \textit{positively regulates}, and \textit{negatively regulates}. These relations describe closely related biological processes and may be difficult to distinguish using graph topology alone, especially when their interpretation depends on biological context or causal direction.

Moreover, 34 of the 42 disagreements have a plausibility score within \(0.05\) of the  decision threshold. 
Rather than indicating a clear model failure, this example illustrates the intrinsic uncertainty of borderline candidate annotations and suggests that future versions of the framework could further exploit adaptive thresholds or curator feedback to refine plausibility evaluation.

\vspace*{-6pt}

\section{Discussion and Conclusions}\label{sec:concl}
In this paper, we addressed the problem of estimating the plausibility of candidate biomedical annotations by exploiting the contextual knowledge encoded in bioKGs. We proposed a relation-specific classification framework based on graph embeddings and a community-based negative sampling strategy for constructing informative training examples. The framework is independent of the underlying graph embedding model and introduces three competition-aware plausibility formulations ({\em Gain}, {\em SoftMax}, and {\em Combo}) that extend conventional confidence estimation by incorporating classifier reliability and competing predicates.

Experiments on five large bioKGs demonstrated the effectiveness of the proposed approach. Community-based negative sampling consistently improved classifier performance over random sampling, yielding more reliable plausibility estimates. Moreover, the proposed formulations better separated target predicates from competing alternatives, enabling a clearer distinction between highly plausible, implausible, and uncertain annotations requiring expert review.

More broadly, our results support the use of bioKGs in combination with machine learning as a decision-support tool for biomedical curation. As the number of candidate annotations exponentially increases, plausibility estimation provides a principled mechanism for prioritizing expert review while preserving human oversight.

The proposed framework also has some limitations that motivate future work. First, plausibility is inferred solely from the structural information encoded in the KG and does not explicitly model semantic relationships among predicates. Incorporating ontology-based knowledge could better distinguish complementary from mutually exclusive relations. Second, the quality of the plausibility estimates depends on the underlying graph representations and relation-specific classifiers. Future work will investigate alternative embedding methods, including graph neural networks, and the integration of textual evidence from the biomedical literature and LLMs. Although evaluated on bioKGs, the framework is general enough to support annotation validation in other heterogeneous KGs.

\vspace*{-9pt}

\section{Data Availability}
Code and data (including the manually annotated dataset for the case study) for reproducing experiments are open-source and available at \url{https://github.com/BioDataUniMI/PlausibilityKG} and \url{https://doi.org/10.5281/zenodo.21359879}.

\vspace*{-9pt}

\section{Competing interests}
There is NO Competing Interest.

\vspace*{-9pt}

\section{Ethics statement}
This study does not involve human participants, animal subjects, or identifiable personal data. All experiments were conducted using publicly available bioKGs. The proposed methodology is intended to support the prioritization of candidate biomedical annotations for expert review and does not substitute expert curation.

\vspace*{-9pt}

\section{Author contributions statement}

M.M. and E.C. conceived the work;  
E.C., D.M., and M.M. developed the plausibility measures;
E.C. and M.A. developed the methods and conducted the experiments;
M.M. and E.C. wrote the paper;
all the authors validated the work.

\vspace*{-6pt}

\section{Funding}

This work was supported by the ``Data Science and Decision Science for Digital Health'' grant under the PSR-2025 funding scheme of the University of Milano.
Support to D.M. has been granted by the National Plan for NRRP Complementary Investments (PNC) in the call for the funding of research initiatives for technologies and innovative trajectories in the health--project n. PNC0000003--AdvaNced Technologies for Human--centrEd Medicine (project acronym: ANTHEM).
Computational resources were provided by INDACO Core facility, which is a project of High Performance Computing at the \href{http://www.unimi.it}{University of Milan}.

\bibliographystyle{plain}
\bibliography{bibliografia}

\end{document}


\maketitle

\setcounter{table}{0}
\renewcommand{\thetable}{S\arabic{table}}
\renewcommand{\tablename}{Supp. Table}

\setcounter{figure}{0}
\renewcommand{\thefigure}{S\arabic{figure}}
\renewcommand{\figurename}{Supp. Fig.}

\renewcommand{\thelstlisting}{S\arabic{lstlisting}}
\renewcommand{\lstlistingname}{Supplementary Listing}

\vspace{-2cm}
\setcounter{section}{0}
\renewcommand{\thesection}{S\arabic{section}}

This Supplementary Material is organized as follows. Sections \ref{sec:appA}–\ref{app:hyperparams} describe datasets and implementation details. Sections \ref{app:details}–\ref{sec:supp_plausibility} provide additional quantitative analyses, while Section \ref{sec:supp_qualitative} presents qualitative examples.

\section{Degree Distributions of bioKGs}
\label{sec:appA}

\begin{wrapfigure}{r}{0.475\linewidth}
    \centering
    \vspace{-1em}
    \includegraphics[width=\linewidth]{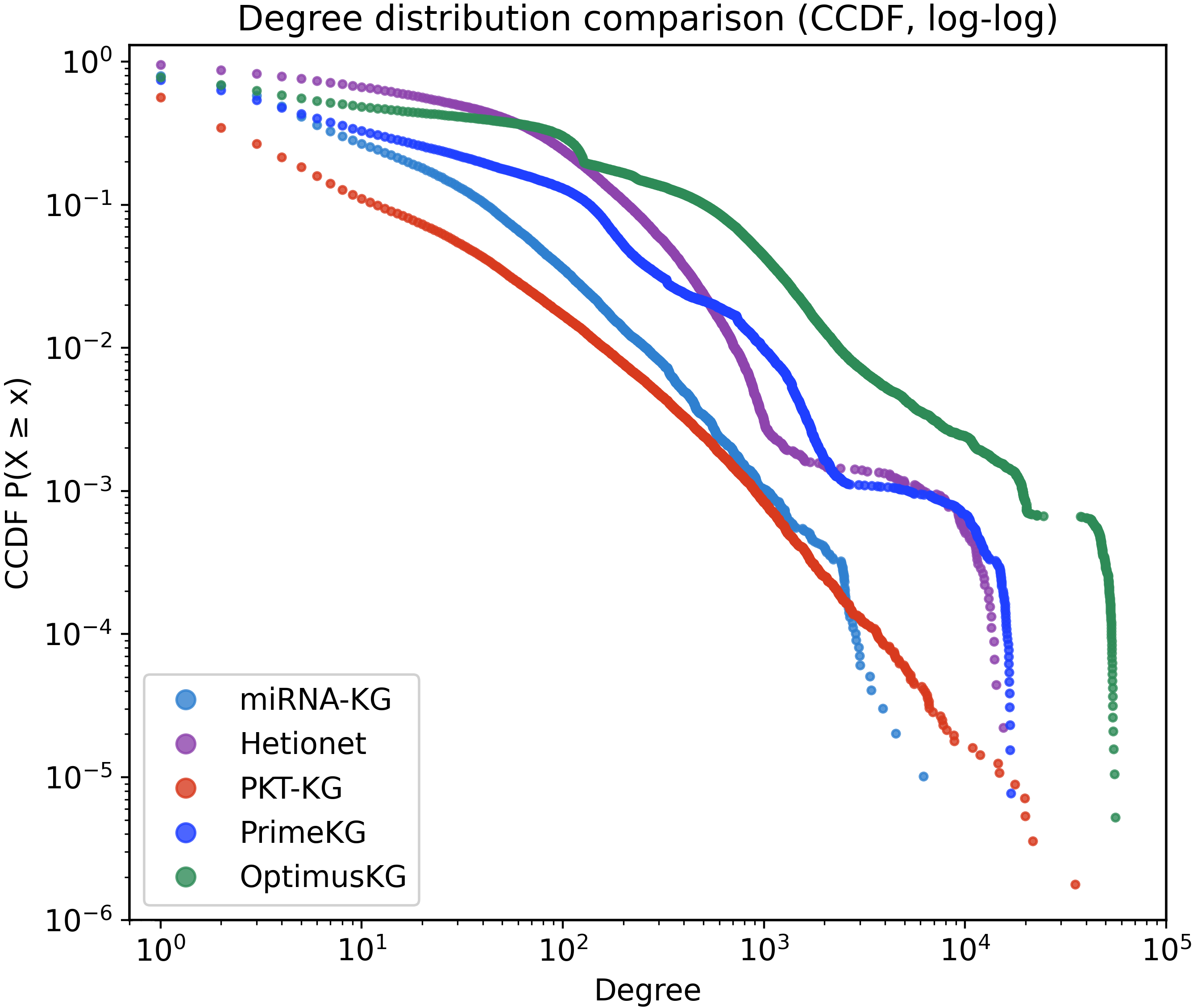}
    \caption{
    Degree distributions of bioKGs.
    }
    \label{fig:ccdf}
    \vspace{-1em}
\end{wrapfigure}

Figure~\ref{fig:ccdf} reports the complementary cumulative degree distributions of the considered bioKGs. 
Across all graphs, a small number of highly connected hub entities coexist with a large number of sparsely connected nodes, revealing strongly heterogeneous connectivity patterns. 
This structural organization is consistent with the presence of densely connected local regions and motivates the use of community-based negative sampling, which exploits graph communities to generate harder and more informative negative examples.

\section{Hyperparameter Configuration}
\label{app:hyperparams}

This appendix reports the hyperparameter grids adopted for the classification models
considered in the evaluation.

For RFs, we tuned the number of estimators and the maximum depth of individual trees.
The number of estimators ranged in 
$\{100, 200, 300, 400, 500\}$, while 
$\it{max\_depth} \in \{\it{None}, 10, 20, 30, 50\}$.

For MLPs, we tuned the regularization strength \(\alpha\) and the network architecture. The regularization parameter was selected from \(13\) logarithmically spaced values in the interval \([10^{-5},10^{-1}]\), while the hidden-layer configurations were chosen from $\{50, 100, 200\}$. Each MLP was trained with early stopping, a tolerance of $10^{-4}$, and a maximum number of $20K$ iterations. 

\section{Detailed Classification Performance}\label{app:details}

Tables~\ref{tab:miRNAKG_top_mean_bottom}–\ref{tab:PrimeKG_top_mean_bottom} detail the classification results obtained for each KG. We report the best five and worst five schema facts in terms of balanced accuracy, together with the negative sampling strategy ({\tt Strategy}), edge cardinality ({\tt Edges}), precision, recall, specificity, $F_1$-score, Matthews correlation coefficient ({\tt MCC}), and $\hat{F}_1$-score.
Results are reported as mean and standard deviation over the outer folds of the nested cross-validation procedure. \

\begin{table*}
\scriptsize
\setlength{\tabcolsep}{2.6725pt}
\begin{tabular}{l p{5.5cm} r ccccccc}
\toprule
Strategy & Schema Fact & Edges & Precision & Recall & Spec. & B. Acc. & $F_{1}$ & MCC & $\hat{F}_{1}$ \\
\midrule
Community & miRNA - in similarity relationship with - miRNA & $1.5K$ & 0.990$\pm$0.004 & 0.978$\pm$0.003 & 0.989$\pm$0.004 & 0.984$\pm$0.003 & 0.984$\pm$0.002 & 0.966$\pm$0.005 & 0.014$\pm$0.003 \\
Random & miRNA - in similarity relationship with - miRNA & $1.5K$ & 0.985$\pm$0.012 & 0.977$\pm$0.014 & 0.987$\pm$0.011 & 0.982$\pm$0.008 & 0.981$\pm$0.008 & 0.964$\pm$0.016 & 0.014$\pm$0.006 \\
Community & Gene - genetically interacts with - Gene & $2.0K$ & 0.967$\pm$0.014 & 0.976$\pm$0.022 & 0.965$\pm$0.015 & 0.970$\pm$0.013 & 0.971$\pm$0.014 & 0.942$\pm$0.027 & 0.024$\pm$0.013 \\
Community & miRNA - regulates activity of - Gene & $36.6K$ & 0.972$\pm$0.002 & 0.975$\pm$0.001 & 0.950$\pm$0.004 & 0.962$\pm$0.002 & 0.974$\pm$0.001 & 0.927$\pm$0.002 & 0.033$\pm$0.001 \\
Community & Gene - causes or contributes to condition - Disease & $36.2K$ & 0.956$\pm$0.003 & 0.955$\pm$0.002 & 0.952$\pm$0.003 & 0.954$\pm$0.002 & 0.956$\pm$0.001 & 0.907$\pm$0.002 & 0.046$\pm$0.001 \\
 &  &  &  &  &  &  &  &  &  \\
Community & {\tt Mean} & $21.3K$ & 0.924$\pm$0.007 & 0.944$\pm$0.010 & 0.886$\pm$0.012 & 0.915$\pm$0.008 & 0.933$\pm$0.006 & 0.839$\pm$0.013 & 0.063$\pm$0.008 \\
Random & {\tt Mean} & $21.3K$ & 0.850$\pm$0.009 & 0.878$\pm$0.012 & 0.857$\pm$0.010 & 0.868$\pm$0.008 & 0.864$\pm$0.008 & 0.736$\pm$0.016 & 0.128$\pm$0.007 \\
 &  &  &  &  &  &  &  &  &  \\
Random & Gene - causes or contributes to condition - Disease & $36.2K$ & 0.860$\pm$0.003 & 0.840$\pm$0.004 & 0.876$\pm$0.003 & 0.858$\pm$0.003 & 0.850$\pm$0.003 & 0.718$\pm$0.005 & 0.139$\pm$0.003 \\
Random & miRNA - has function - GO & $4.4K$ & 0.792$\pm$0.012 & 0.829$\pm$0.015 & 0.804$\pm$0.014 & 0.817$\pm$0.010 & 0.810$\pm$0.011 & 0.633$\pm$0.020 & 0.182$\pm$0.010 \\
Random & miRNA - regulates activity of - Gene & $36.6K$ & 0.761$\pm$0.008 & 0.829$\pm$0.005 & 0.766$\pm$0.010 & 0.797$\pm$0.006 & 0.794$\pm$0.005 & 0.594$\pm$0.011 & 0.198$\pm$0.005 \\
Community & miRNA - is causal somatic mutation in - Disease & $1.4K$ & 0.772$\pm$0.009 & 0.938$\pm$0.023 & 0.657$\pm$0.023 & 0.797$\pm$0.016 & 0.847$\pm$0.008 & 0.631$\pm$0.019 & 0.104$\pm$0.033 \\
Random & miRNA - is causal somatic mutation in - Disease & $1.4K$ & 0.655$\pm$0.015 & 0.747$\pm$0.021 & 0.645$\pm$0.023 & 0.696$\pm$0.014 & 0.698$\pm$0.014 & 0.393$\pm$0.028 & 0.294$\pm$0.015 \\
\bottomrule
\end{tabular}\caption{Detailed performance on miRNA-KG.}\label{tab:miRNAKG_top_mean_bottom}
\end{table*}

\begin{table*}
\scriptsize
\setlength{\tabcolsep}{2.6725pt}
\begin{tabular}{l p{5.5cm} r ccccccc}
\toprule
Strategy & Schema Fact & Edges & Precision & Recall & Spec. & B. Acc. & $F_1$ & MCC & $\hat{F}_{1}$ \\
\midrule
Community & Anatomy - expresses - Gene & $526.4K$ & 0.988$\pm$0.000 & 0.996$\pm$0.000 & 0.935$\pm$0.001 & 0.966$\pm$0.001 & 0.992$\pm$0.000 & 0.949$\pm$0.001 & 0.007$\pm$0.000 \\
Community & Gene - participates - Cellular component & $73.6K$ & 0.911$\pm$0.001 & 0.956$\pm$0.002 & 0.858$\pm$0.001 & 0.907$\pm$0.000 & 0.933$\pm$0.001 & 0.827$\pm$0.001 & 0.067$\pm$0.002 \\
Community & Gene - covaries - Gene & $61.7K$ & 0.926$\pm$0.002 & 0.888$\pm$0.002 & 0.925$\pm$0.002 & 0.906$\pm$0.002 & 0.906$\pm$0.002 & 0.812$\pm$0.004 & 0.090$\pm$0.002 \\
Random & Anatomy - expresses - Gene & $526.4K$ & 0.803$\pm$0.002 & 0.976$\pm$0.000 & 0.785$\pm$0.003 & 0.881$\pm$0.002 & 0.881$\pm$0.001 & 0.769$\pm$0.003 & 0.044$\pm$0.000 \\
Community & Gene - interacts - Gene & $147.2K$ & 0.894$\pm$0.002 & 0.897$\pm$0.002 & 0.857$\pm$0.003 & 0.877$\pm$0.001 & 0.895$\pm$0.001 & 0.754$\pm$0.002 & 0.120$\pm$0.001 \\
 &  &  &  &  &  &  &  &  &  \\
Community & {\tt Mean} & $107.1K$ & 0.819$\pm$0.007 & 0.855$\pm$0.007 & 0.794$\pm$0.011 & 0.824$\pm$0.006 & 0.835$\pm$0.005 & 0.656$\pm$0.011 & 0.153$\pm$0.006 \\
Random & {\tt Mean} & $107.1K$ & 0.734$\pm$0.006 & 0.765$\pm$0.009 & 0.744$\pm$0.008 & 0.754$\pm$0.006 & 0.746$\pm$0.006 & 0.512$\pm$0.011 & 0.228$\pm$0.007 \\
 &  &  &  &  &  &  &  &  &  \\
Community & Anatomy - upregulates - Gene & $97.8K$ & 0.622$\pm$0.018 & 0.780$\pm$0.004 & 0.489$\pm$0.043 & 0.634$\pm$0.019 & 0.692$\pm$0.010 & 0.281$\pm$0.037 & 0.307$\pm$0.004 \\
Random & Disease - upregulates - Gene & $7.7K$ & 0.572$\pm$0.006 & 0.808$\pm$0.014 & 0.456$\pm$0.010 & 0.632$\pm$0.008 & 0.670$\pm$0.008 & 0.280$\pm$0.019 & 0.284$\pm$0.015 \\
Random & Disease - downregulates - Gene & $7.6K$ & 0.561$\pm$0.005 & 0.808$\pm$0.011 & 0.432$\pm$0.006 & 0.620$\pm$0.007 & 0.662$\pm$0.006 & 0.257$\pm$0.015 & 0.287$\pm$0.012 \\
Random & Anatomy - upregulates - Gene & $97.8K$ & 0.596$\pm$0.002 & 0.610$\pm$0.004 & 0.628$\pm$0.003 & 0.619$\pm$0.002 & 0.603$\pm$0.003 & 0.238$\pm$0.004 & 0.381$\pm$0.002 \\
Random & Disease - localizes - Anatomy & $3.6K$ & 0.622$\pm$0.024 & 0.520$\pm$0.028 & 0.715$\pm$0.033 & 0.617$\pm$0.015 & 0.566$\pm$0.018 & 0.240$\pm$0.032 & 0.356$\pm$0.025 \\
\bottomrule
\end{tabular}\caption{Detailed performance on Hetionet.}\label{tab:hetionet_top_mean_bottom}
\end{table*}

\begin{table*}
\scriptsize
\setlength{\tabcolsep}{2.6725pt}
\begin{tabular}{l p{5.5cm} r ccccccc}
\toprule
Strategy & Schema Fact & Edges & Precision & Recall & Spec. & B. Acc. & $F_1$ & MCC & $\hat{F}_{1}$ \\
\midrule
Community & Gene - interacts with - Protein & $142.1K$ & 0.988$\pm$0.001 & 0.983$\pm$0.001 & 0.974$\pm$0.002 & 0.978$\pm$0.001 & 0.985$\pm$0.000 & 0.954$\pm$0.001 & 0.021$\pm$0.001 \\
Community & Gene - genetically interacts with - Gene & $2.0K$ & 0.974$\pm$0.008 & 0.975$\pm$0.008 & 0.972$\pm$0.008 & 0.974$\pm$0.006 & 0.974$\pm$0.002 & 0.947$\pm$0.004 & 0.024$\pm$0.003 \\
Community & Protein - molecularly interacts with - Protein & $341.1K$ & 0.968$\pm$0.001 & 0.985$\pm$0.001 & 0.948$\pm$0.001 & 0.967$\pm$0.001 & 0.976$\pm$0.001 & 0.938$\pm$0.001 & 0.024$\pm$0.001 \\
Community & Gene - causes or contributes to condition - Disease & $36.2K$ & 0.965$\pm$0.003 & 0.955$\pm$0.003 & 0.962$\pm$0.004 & 0.958$\pm$0.003 & 0.960$\pm$0.003 & 0.917$\pm$0.005 & 0.041$\pm$0.003 \\
Random & Gene - genetically interacts with - Gene & $2.0K$ & 0.947$\pm$0.017 & 0.960$\pm$0.013 & 0.952$\pm$0.016 & 0.956$\pm$0.009 & 0.954$\pm$0.009 & 0.912$\pm$0.018 & 0.041$\pm$0.009 \\
 &  &  &  &  &  &  &  &  &  \\
Community & {\tt Mean} & $86.4K$ & 0.863$\pm$0.006 & 0.900$\pm$0.006 & 0.846$\pm$0.008 & 0.873$\pm$0.006 & 0.879$\pm$0.005 & 0.750$\pm$0.010 & 0.108$\pm$0.005 \\
Random & {\tt Mean} & $86.4K$ & 0.824$\pm$0.006 & 0.850$\pm$0.007 & 0.832$\pm$0.008 & 0.841$\pm$0.005 & 0.835$\pm$0.005 & 0.685$\pm$0.009 & 0.144$\pm$0.005 \\
 &  &  &  &  &  &  &  &  &  \\
Community & GO - positively regulates - GO & $2.5K$ & 0.711$\pm$0.015 & 0.930$\pm$0.022 & 0.660$\pm$0.018 & 0.795$\pm$0.014 & 0.806$\pm$0.017 & 0.607$\pm$0.039 & 0.115$\pm$0.031 \\
Community & Protein - located in - Anatomy & $16.1K$ & 0.698$\pm$0.006 & 0.696$\pm$0.012 & 0.715$\pm$0.005 & 0.706$\pm$0.007 & 0.697$\pm$0.009 & 0.410$\pm$0.015 & 0.295$\pm$0.007 \\
Community & Protein - located in - Cell & $10.8K$ & 0.690$\pm$0.009 & 0.708$\pm$0.005 & 0.680$\pm$0.014 & 0.694$\pm$0.007 & 0.699$\pm$0.004 & 0.389$\pm$0.012 & 0.305$\pm$0.006 \\
Random & Protein - located in - Anatomy & $16.1K$ & 0.646$\pm$0.005 & 0.673$\pm$0.020 & 0.668$\pm$0.005 & 0.671$\pm$0.008 & 0.659$\pm$0.012 & 0.341$\pm$0.017 & 0.329$\pm$0.009 \\
Random & Protein - located in - Cell & $10.8K$ & 0.617$\pm$0.005 & 0.597$\pm$0.009 & 0.666$\pm$0.003 & 0.632$\pm$0.006 & 0.607$\pm$0.007 & 0.264$\pm$0.011 & 0.365$\pm$0.005 \\
\bottomrule
\end{tabular}\caption{Detailed performance on PKT-KG.}\label{tab:PKTKG_top_mean_bottom}
\end{table*}

\begin{table*}
\scriptsize
\setlength{\tabcolsep}{2.6725pt}
\begin{tabular}{l p{5.5cm} r ccccccc}
\toprule
Strategy & Schema Fact & Edges & Precision & Recall & Spec. & B. Acc. & $F_1$ & MCC & $\hat{F}_{1}$ \\
\midrule
Community & Disease - phenotype present - Effect and/or phenotype & $150.3K$ & 0.973$\pm$0.001 & 0.969$\pm$0.001 & 0.928$\pm$0.003 & 0.948$\pm$0.002 & 0.971$\pm$0.001 & 0.893$\pm$0.003 & 0.043$\pm$0.001 \\
Community & Gene and/or protein - interacts with - Molecular function & $69.5K$ & 0.944$\pm$0.001 & 0.941$\pm$0.002 & 0.917$\pm$0.001 & 0.929$\pm$0.002 & 0.942$\pm$0.002 & 0.856$\pm$0.004 & 0.070$\pm$0.002 \\
Community & Drug - enzyme - Gene and/or protein & $5.3K$ & 0.971$\pm$0.003 & 0.884$\pm$0.013 & 0.971$\pm$0.003 & 0.927$\pm$0.008 & 0.925$\pm$0.008 & 0.855$\pm$0.015 & 0.046$\pm$0.004 \\
Community & Cellular component - interacts with - Gene and/or protein & $83.4K$ & 0.922$\pm$0.002 & 0.940$\pm$0.002 & 0.890$\pm$0.004 & 0.915$\pm$0.002 & 0.931$\pm$0.001 & 0.834$\pm$0.003 & 0.077$\pm$0.002 \\
Random & Anatomy - expression present - Gene and/or protein & $1.5M$ & 0.889$\pm$0.001 & 0.916$\pm$0.001 & 0.897$\pm$0.001 & 0.907$\pm$0.001 & 0.902$\pm$0.000 & 0.812$\pm$0.000 & 0.092$\pm$0.000 \\
 &  &  &  &  &  &  &  &  &  \\
Community & {\tt Mean} & $253.0K$ & 0.874$\pm$0.007 & 0.876$\pm$0.010 & 0.824$\pm$0.010 & 0.850$\pm$0.006 & 0.873$\pm$0.006 & 0.715$\pm$0.011 & 0.113$\pm$0.006 \\
Random & {\tt Mean} & $253.0K$ & 0.789$\pm$0.009 & 0.753$\pm$0.013 & 0.819$\pm$0.009 & 0.786$\pm$0.007 & 0.768$\pm$0.009 & 0.576$\pm$0.014 & 0.198$\pm$0.007 \\
 &  &  &  &  &  &  &  &  &  \\
Community & Anatomy - expression absent - Gene and/or protein & $19.9K$ & 0.756$\pm$0.003 & 0.933$\pm$0.003 & 0.444$\pm$0.010 & 0.689$\pm$0.004 & 0.835$\pm$0.002 & 0.451$\pm$0.007 & 0.119$\pm$0.004 \\
Random & Anatomy - expression absent - Gene and/or protein & $19.9K$ & 0.674$\pm$0.007 & 0.666$\pm$0.013 & 0.710$\pm$0.011 & 0.688$\pm$0.006 & 0.670$\pm$0.008 & 0.376$\pm$0.013 & 0.310$\pm$0.006 \\
Community & Disease - phenotype absent - Effect and/or phenotype & $1.2K$ & 0.651$\pm$0.018 & 0.516$\pm$0.054 & 0.748$\pm$0.042 & 0.632$\pm$0.012 & 0.574$\pm$0.031 & 0.273$\pm$0.021 & 0.327$\pm$0.024 \\
Random & Disease - off label use - Drug & $2.6K$ & 0.606$\pm$0.016 & 0.578$\pm$0.019 & 0.661$\pm$0.016 & 0.620$\pm$0.014 & 0.591$\pm$0.016 & 0.240$\pm$0.028 & 0.376$\pm$0.014 \\
Random & Disease - phenotype absent - Effect and/or phenotype & $1.2K$ & 0.562$\pm$0.030 & 0.466$\pm$0.057 & 0.674$\pm$0.035 & 0.570$\pm$0.027 & 0.508$\pm$0.043 & 0.143$\pm$0.054 & 0.403$\pm$0.027 \\
\bottomrule
\end{tabular}\caption{Detailed performance on PrimeKG.}\label{tab:PrimeKG_top_mean_bottom}
\end{table*}

\begin{table*}
\scriptsize
\setlength{\tabcolsep}{2.6725pt}
\begin{tabular}{l p{5.5cm} r ccccccc}
\toprule
Strategy & Schema Fact & Edges & Precision & Recall & Spec. & B. Acc. & $F_1$ & MCC & $\hat{F}_{1}$ \\
\midrule
Community & Anatomy - EXPRESSION PRESENT - Gene & $6.6M$ & 0.995$\pm$0.000 & 1.000$\pm$0.000 & 0.950$\pm$0.001 & 0.975$\pm$0.000 & 0.997$\pm$0.000 & 0.970$\pm$0.001 & 0.001$\pm$0.000 \\
Community & Anatomy - EXPRESSION ABSENT - Gene & $2.2M$ & 0.962$\pm$0.000 & 0.946$\pm$0.001 & 0.961$\pm$0.000 & 0.953$\pm$0.000 & 0.954$\pm$0.000 & 0.906$\pm$0.000 & 0.045$\pm$0.001 \\
Community & Disease - ASSOCIATED WITH - Gene & $9.7M$ & 0.954$\pm$0.000 & 0.987$\pm$0.000 & 0.895$\pm$0.001 & 0.941$\pm$0.000 & 0.970$\pm$0.000 & 0.902$\pm$0.000 & 0.024$\pm$0.000 \\
Community & Cellular component - INTERACTS WITH - Gene & $105.3K$ & 0.915$\pm$0.002 & 0.922$\pm$0.002 & 0.897$\pm$0.003 & 0.909$\pm$0.001 & 0.918$\pm$0.001 & 0.820$\pm$0.003 & 0.089$\pm$0.001 \\
Community & Molecular function - INTERACTS WITH - Gene & $90.9K$ & 0.931$\pm$0.001 & 0.890$\pm$0.003 & 0.923$\pm$0.002 & 0.907$\pm$0.002 & 0.910$\pm$0.002 & 0.811$\pm$0.003 & 0.091$\pm$0.001 \\
 &  &  &  &  &  &  &  &  &  \\
Community & {\tt Mean} & $941.4K$ & 0.855$\pm$0.008 & 0.885$\pm$0.009 & 0.804$\pm$0.011 & 0.844$\pm$0.008 & 0.868$\pm$0.007 & 0.701$\pm$0.015 & 0.125$\pm$0.008 \\
Random & {\tt Mean} & $941.4K$ & 0.757$\pm$0.008 & 0.759$\pm$0.013 & 0.779$\pm$0.009 & 0.769$\pm$0.007 & 0.756$\pm$0.009 & 0.541$\pm$0.014 & 0.217$\pm$0.007 \\
 &  &  &  &  &  &  &  &  &  \\
Community & Drug - SYNERGISTIC INTERACTION - Drug & $1.3M$ & 0.667$\pm$0.001 & 0.962$\pm$0.001 & 0.404$\pm$0.002 & 0.683$\pm$0.001 & 0.788$\pm$0.001 & 0.453$\pm$0.001 & 0.072$\pm$0.001 \\
Community & Drug - OFF LABEL USE - Disease & $1.1K$ & 0.663$\pm$0.036 & 0.659$\pm$0.038 & 0.689$\pm$0.033 & 0.674$\pm$0.035 & 0.661$\pm$0.036 & 0.347$\pm$0.069 & 0.325$\pm$0.034 \\
Random & Drug - ASSOCIATED WITH - Phenotype & $3.7K$ & 0.686$\pm$0.014 & 0.573$\pm$0.032 & 0.765$\pm$0.006 & 0.669$\pm$0.017 & 0.624$\pm$0.025 & 0.345$\pm$0.032 & 0.303$\pm$0.009 \\
Random & Drug - INDICATION - Phenotype & $1.0K$ & 0.595$\pm$0.032 & 0.542$\pm$0.026 & 0.668$\pm$0.037 & 0.605$\pm$0.024 & 0.567$\pm$0.025 & 0.211$\pm$0.050 & 0.384$\pm$0.029 \\
Random & Drug - OFF LABEL USE - Disease & $1.1K$ & 0.526$\pm$0.018 & 0.506$\pm$0.052 & 0.591$\pm$0.027 & 0.549$\pm$0.020 & 0.515$\pm$0.035 & 0.097$\pm$0.039 & 0.445$\pm$0.015 \\
\bottomrule
\end{tabular}\caption{Detailed performance on OptimusKG.}\label{tab:OptimusKG_top_mean_bottom}
\end{table*}

\section{Statistical Validation of the Classification Results}
\label{sec:supp_statistical}

\begin{table}[t]
\centering
\small
\setlength{\tabcolsep}{3pt}

\begin{minipage}{0.46\textwidth}
\centering
\subfloat[\label{tab:plausibility_soft}]{
\begin{tabular}{lc|cc}
\toprule
Metric & \(k=2\) & \(k=1\) & \(k=3\) \\
\midrule
PAR & 63.0\% (+3.7) & \textbf{64.7\% (+3.3)} & 62.9\% (+3.7) \\
NRR & 82.8\% (+3.9) & 80.4\% (+4.5) & \textbf{83.1\% (+3.9)} \\
Cal & 72.9\% (+3.8) & 72.5\% (+3.9) & \textbf{73.0\% (+3.8)} \\
Sep & 48.0\% (+3.8) & \textbf{49.7\% (+3.6)} & 47.7\% (+3.9) \\
Sep$^{best}$ & 94.3\% (+1.2) & \textbf{99.3\% ($\pm$0.0)} & 93.8\% (+1.3) \\
Sep$^{worst}$ & 72.3\% (+0.8) & 69.4\% (+0.8) & \textbf{72.6\% (+0.5)} \\
\bottomrule
\end{tabular}
}
\end{minipage}
\hfill
\begin{minipage}{0.49\textwidth}
\centering
\subfloat[\label{tab:plausibility_combo}]{
\begin{tabular}{lc|cc}
\toprule
Metric & \(\alpha=0.5\) & \(\alpha=0.25\) & \(\alpha=0.75\) \\
\midrule
PAR & 76.9\% (+4.6) & 75.1\% (+1.8) & \textbf{78.8\% (+7.4)} \\
NRR & 69.7\% (+3.9) & 61.2\% (+4.7) & \textbf{78.1\% (+3.2)} \\
Cal & 73.3\% (+4.3) & 68.1\% (+3.2) & \textbf{78.5\% (+5.3)} \\
Sep & 37.2\% (+2.2) & \textbf{42.9\% (+3.0)} & 32.7\% (+1.2) \\
Sep$^{best}$ & 84.9\% (+6.7) & \textbf{94.5\% (+2.2)} & 67.0\% (+4.6) \\
Sep$^{worst}$ & \textbf{77.7\% (+2.6)} & 76.1\% (+1.9) & 70.7\% (+0.6) \\
\bottomrule
\end{tabular}
}
\end{minipage}

\caption{Evaluation of parameter settings for (a) \(\mathcal{P}_{\tt soft}\) and (b) \(\mathcal{P}_{\tt combo}\) across all schema facts. Best value for each metric is highlighted in bold.}
\label{tab:plausibility_details}
\end{table}

\subsection{Comparison of prediction distributions}

The community-based and random negative sampling strategies are trained using different sets of generated negative triples. Consequently, the predictions produced by the two classifiers cannot be directly paired over the complete evaluation datasets because each model is exposed to a different collection of negative examples.

To assess whether the two strategies exhibit different prediction behaviors, we applied Pearson's chi-square test of independence to the binary predictions obtained over the blind positive triples \(BlP_i\) and the corresponding generated negatives \(N_i\) for each schema fact.
The null hypothesis assumes that the predictions produced by the two strategies are statistically independent. Rejecting this hypothesis indicates that the two sampling strategies induce significantly different prediction distributions.
Across the 98 evaluated schema facts, the null hypothesis was rejected (\(p<0.001\)) in 60 cases, demonstrating a strong statistical association between the predictions produced by the two training strategies. 

However, the \(BlP_i\) samples are the same for both strategies. We therefore applied McNemar's test on the binary predictions obtained for \(BlP_i\) to assess whether the two strategies differ in their predictions on the same plausible instances. In this case, a small \(p\)-value has a different meaning: it indicates that the disagreements are significantly unbalanced toward one strategy. The test reveals a significant asymmetry (\(p < 0.001\) in \(85\) out of \(98\) schema facts).

\subsection{Comparison of classifier confidence}

Since McNemar's test only indicates that the discordant predictions are unbalanced, we further compared the median scores assigned by the two strategies on the same \(BlP_i\) samples to identify the strategy providing higher confidence to true positives. 


For each schema fact, we computed the median classifier score over the corresponding blind test set. Median values were preferred over means because they are less affected by the presence of a small number of extremely confident predictions.
The community-based strategy assigns higher median confidence than the random baseline for 92 out of the 98 schema facts (\(93.9\%\)). The average improvement is \(10.13\) percentage points, while the maximum observed increase reaches \(31.80\) points for the schema fact \textit{(Anatomy,expression-present,Gene)} in OptimusKG.
Moreover, the improvement exceeds five percentage points for 64 schema facts (\(65.3\%\)) and ten percentage points for 45 schema facts (\(45.9\%\)). These results indicate that community-based negative sampling systematically increases the confidence assigned to previously unseen plausible facts
.


\section{Additional Quantitative Evaluation of Plausibility}
\label{sec:supp_plausibility}

\subsection{Sensitivity to Parameter Selection}

The plausibility formulations introduce a small number of parameters controlling the contribution of classifier reliability and competition-aware information. To assess the robustness of the proposed approach, we evaluated different parameter configurations
.

For the Gain and SoftMax formulations, the scaling parameter $\lambda$ was varied between 1 and 20. Small values produce smoother plausibility distributions, whereas larger values increasingly emphasize the differences between competing predicates. Performance remains stable
, with $\lambda=10$ providing the best compromise between calibration and predicate separation.

For the Combo formulation, we evaluated the mixing parameter $\alpha$ between 0 and 1. 
Intermediate values provide the best trade-off, with $\alpha=0.5$ yielding the highest average performance across the proposed plausibility metrics.

For the SoftMax formulation, we also investigated the number $k$ of competing predicates included in the normalization. Small values focus on the strongest alternatives, whereas larger values incorporate increasingly weaker competitors. We observed limited improvements beyond $k=2$, indicating that the strongest competitors already capture most of the relevant competition among predicates.

Table~\ref{tab:plausibility_details} reports additional parameter configurations for \(\mathcal{P}_{\tt soft}\) and \(\mathcal{P}_{\tt combo}\). Values are percentages; values in parentheses report the improvement, in percentage points, obtained by using community-based negatives instead of the random baseline.



\subsection{Impact of Community-Based Negative Sampling}

The main paper reports the aggregate improvements obtained by replacing random negatives with the proposed community-based strategy. Here we provide a more detailed analysis.

Across the 24 formulation--metric combinations considered in Table 5
, community-based negatives improve performance in 22 cases, with an average gain of approximately three percentage points.
The largest improvement is observed for the Base formulation on the Positive Acceptance Rate, where the increase exceeds ten percentage points. Since the Base formulation directly reflects the confidence assigned by the underlying classifier, this result confirms that community-aware negatives primarily improve the classifier's ability to recognize previously unseen plausible triples.

Competition-aware formulations also consistently benefit from the improved classifiers. Although the gains are smaller than those observed for the Base formulation, they are remarkably stable across the separation metrics, demonstrating that better classifier calibration naturally translates into more informative competition-aware plausibility estimates.

\subsection{Analysis of Plausibility Score Polarization}

Besides calibration and competition-aware discrimination, we also investigated how the different plausibility formulations distribute scores over the interval $[0,1]$.

For each schema fact, we considered the plausibility scores assigned to blind plausible triples $BlP_i$ and generated negatives $N_i$, obtaining the empirical score distribution $X$. We quantify score polarization through the normalized dispersion index

$D
=
\frac{\mathrm{Var}(X)}
{\mathrm{Mean}(X)\left(1-\mathrm{Mean}(X)\right)}.$

The denominator corresponds to the maximum variance attainable by a distribution on $[0,1]$ having the same mean. Consequently, $D$ assumes values in $[0,1]$, where larger values indicate increasingly polarized score distributions.

A continuous uniform distribution on $[0,1]$ has a dispersion index $D=1/3$. Accordingly, we consider values exceeding this threshold as evidence of score polarization.
Using community-based negatives, the Combo formulation satisfies this criterion in 74 out of 98 schema facts (75.5\%), compared with 64 (65.3\%) when random negatives are adopted.
The same behavior is observed for all formulations. Base satisfies the criterion in 90 schema facts using community-based negatives, compared with 79 using random negatives. Gain increases from 59 to 66 schema facts, whereas SoftMax increases from 58 to 65.
The difference becomes even more evident for highly polarized distributions. Considering all formulations together, the stricter criterion $D>2/3$ is satisfied in 132 cases using community-based negatives, but in only 10 cases using random negatives. Similarly, $D>0.8$ is observed in 63 cases with community-based negatives, compared with only three cases for random sampling.




\begin{figure}[t]
    \centering

    \begin{minipage}{0.46\linewidth}
        \centering
        \subfloat[Community-based negatives
        \label{fig:bio_competitors_upregulates_comm}]{
            \includegraphics[width=\linewidth]{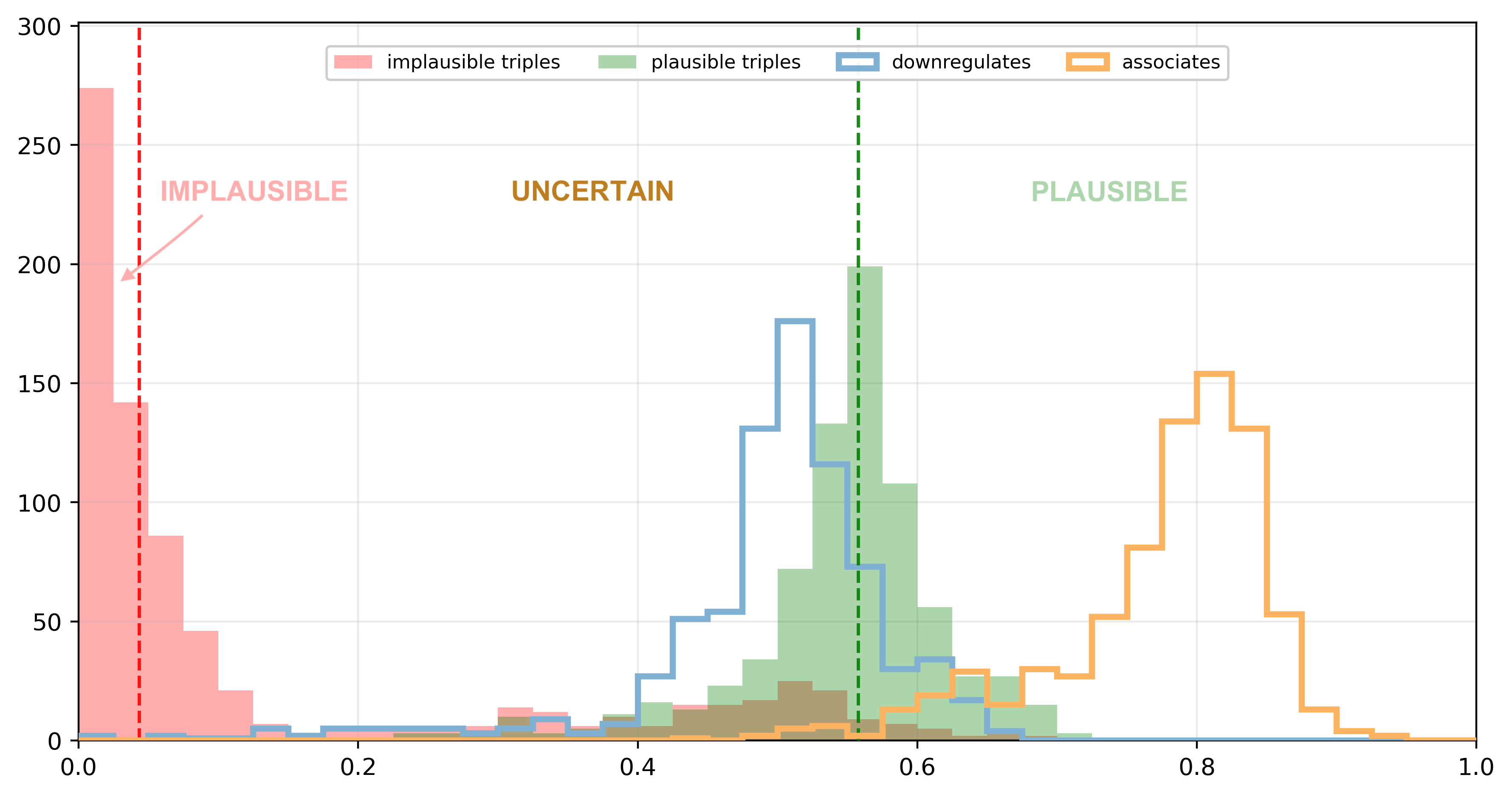}
        }
    \end{minipage}
    \hfill
    \begin{minipage}{0.46\linewidth}
        \centering
        \subfloat[Random negatives
        \label{fig:bio_competitors_upregulates_rand}]{
            \includegraphics[width=\linewidth]{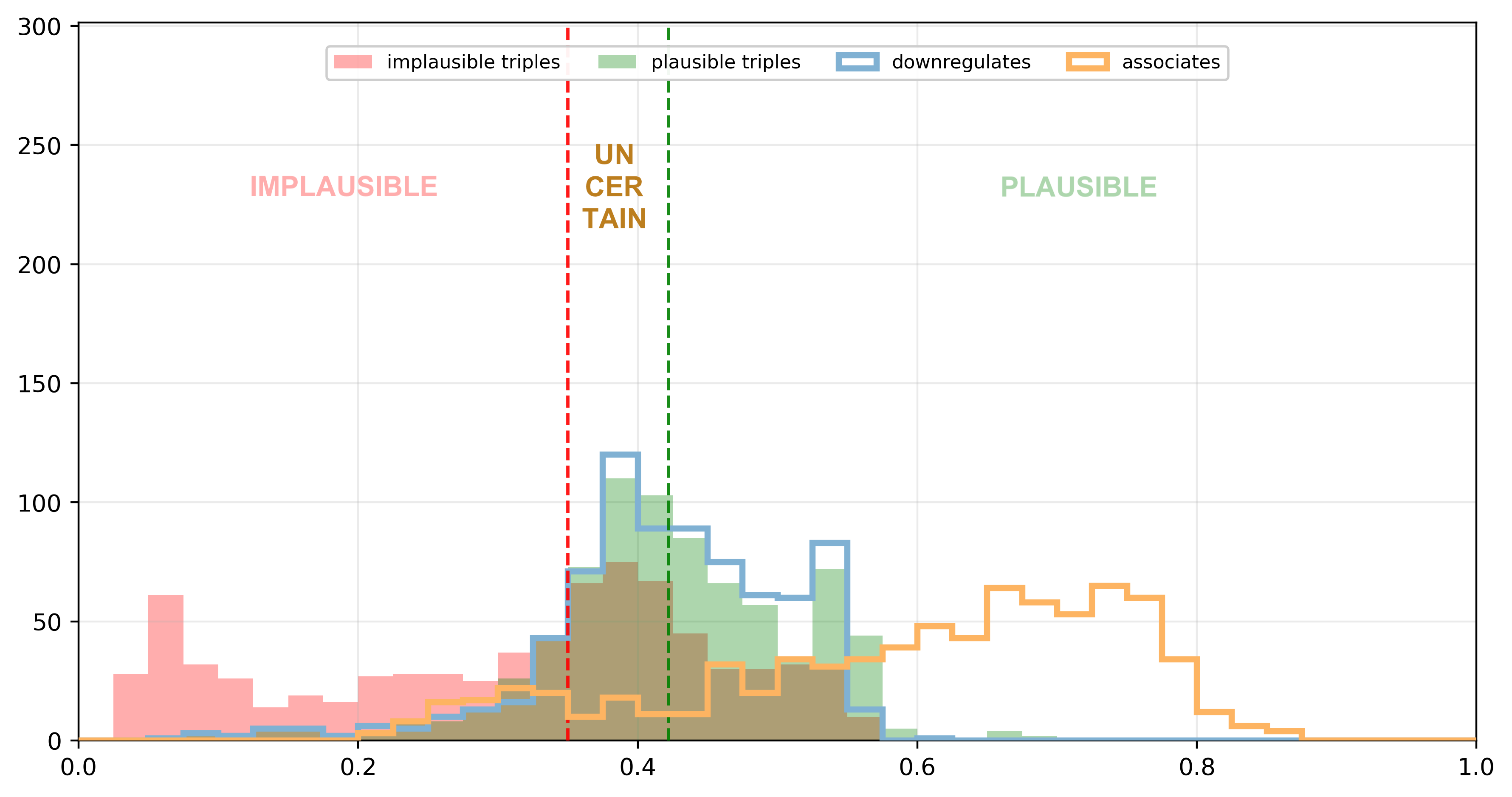}
        }
    \end{minipage}

    \caption{
    Distribution of \(\mathcal{P}_{\tt combo}\) scores for the schema fact
    \textit{(Disease,upregulates,Gene)} in Hetionet.
    }
    \label{fig:bio_competitors_upregulates}
\end{figure}

\begin{figure}[t]
    \centering

    \begin{minipage}{0.46\linewidth}
        \centering
        \subfloat[Community-based negatives]{
            \includegraphics[width=\linewidth]{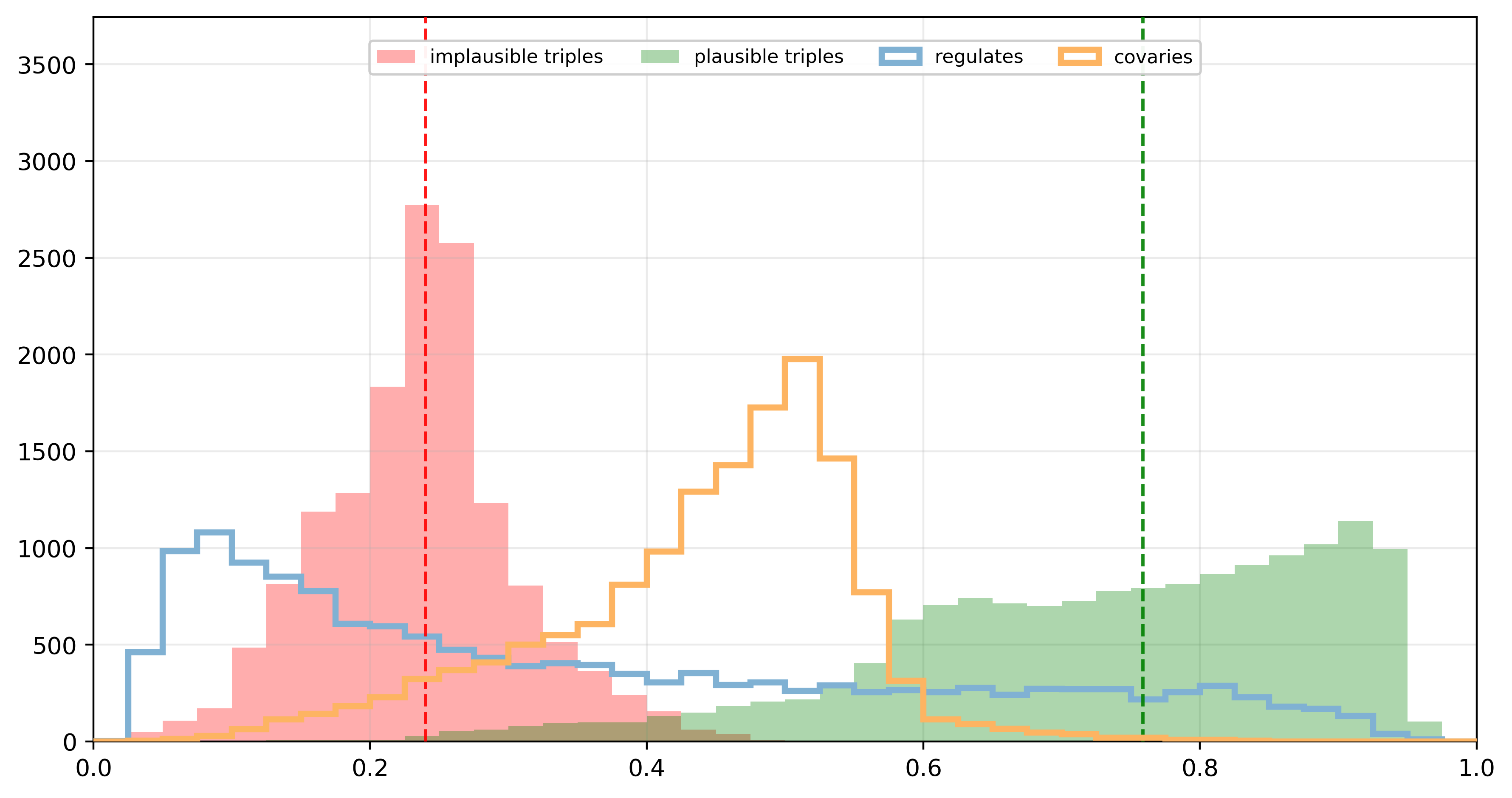}
        }
    \end{minipage}
    \hfill
    \begin{minipage}{0.46\linewidth}
        \centering
        \subfloat[Random negatives]{
            \includegraphics[width=\linewidth]{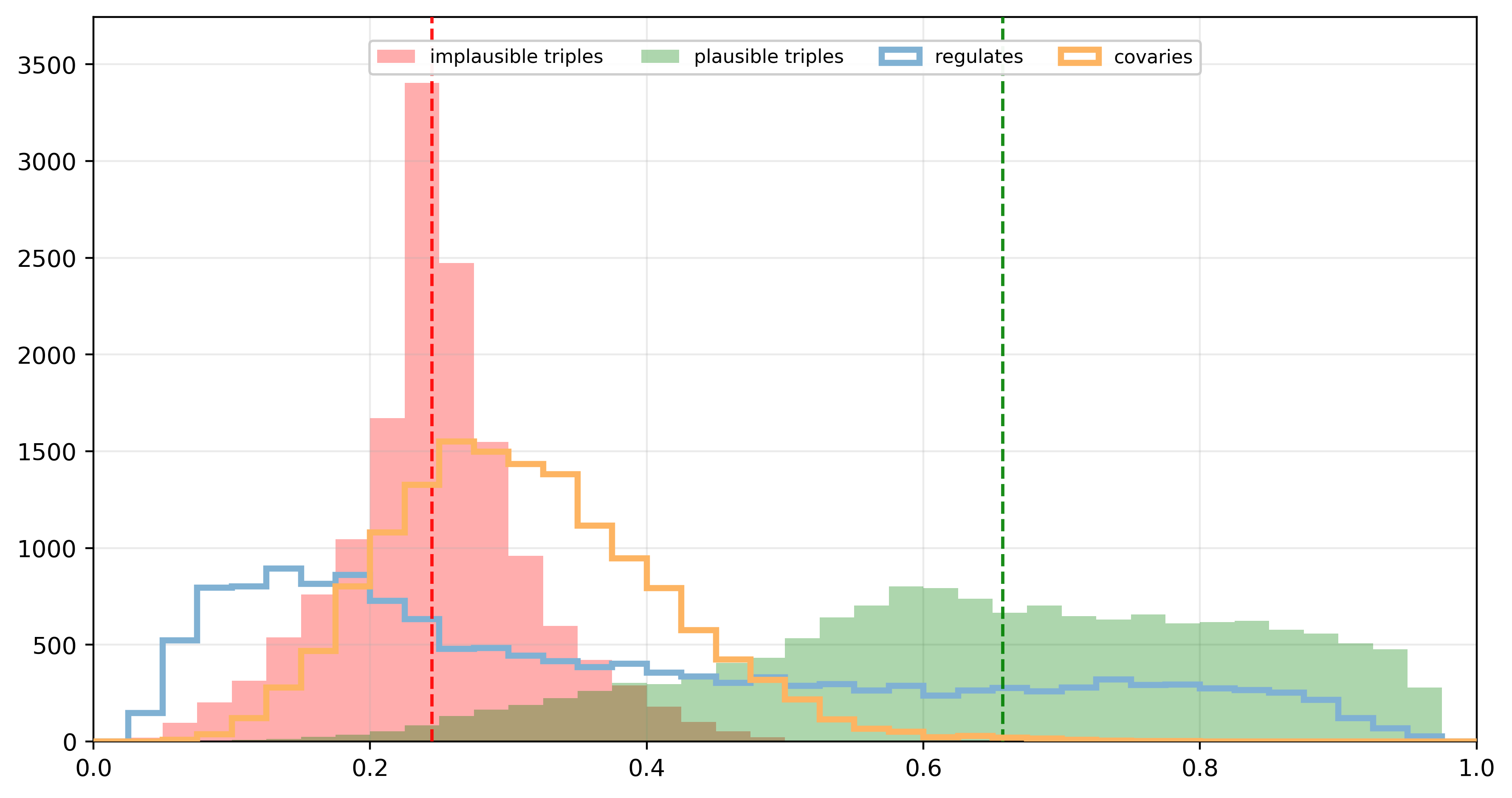}
        }
    \end{minipage}

    \caption{
    Distribution of \(\mathcal{P}_{\tt combo}\) scores for the schema fact \textit{(Gene,interacts,Gene)} in Hetionet.
    }
    \label{fig:app_gene_gene_interacts}
\end{figure}

\begin{figure}[t]
    \centering

    \begin{minipage}{0.46\linewidth}
        \centering
        \subfloat[Community-based negatives]{
            \includegraphics[width=\linewidth]{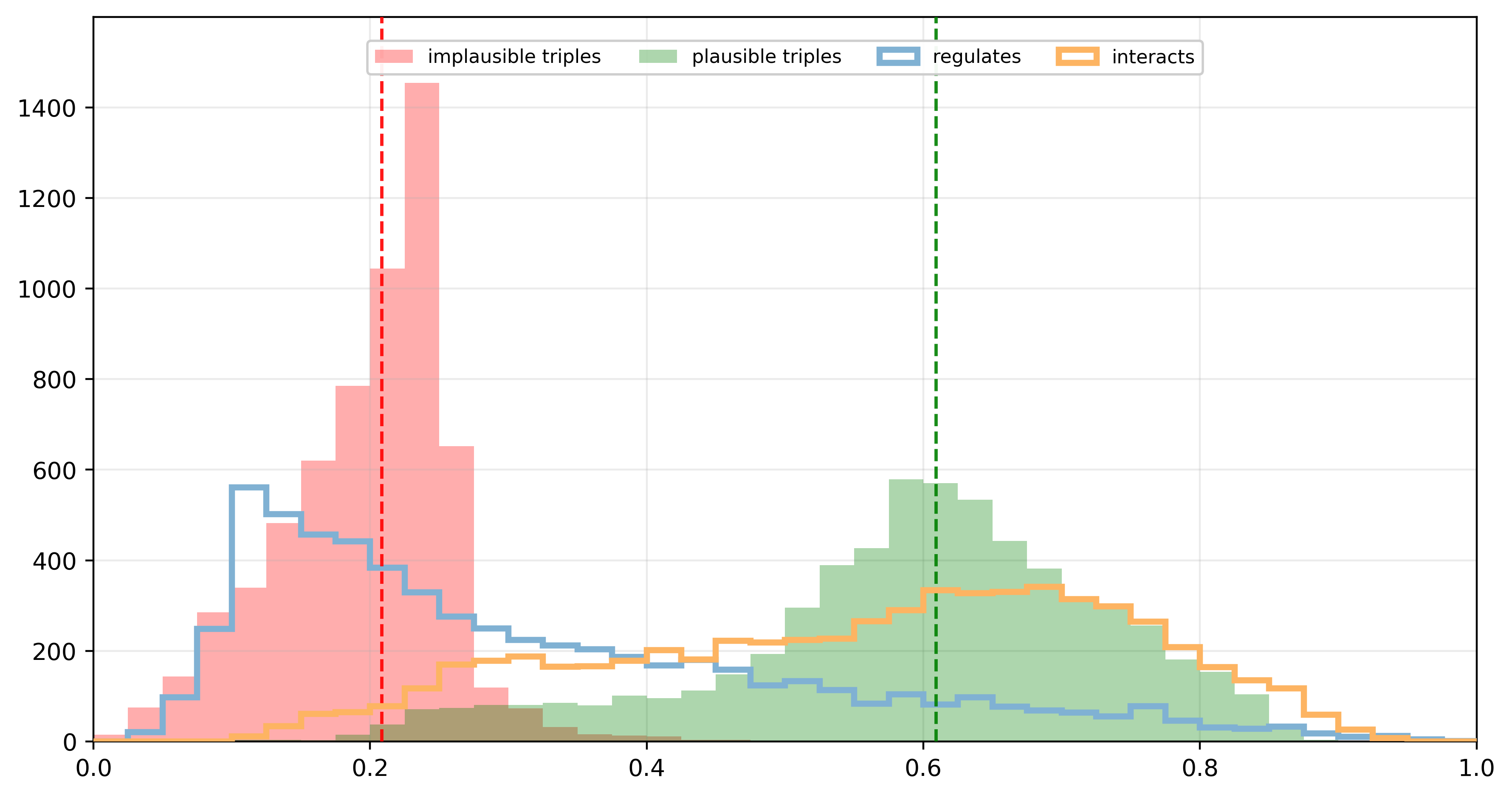}
        }
    \end{minipage}
    \hfill
    \begin{minipage}{0.46\linewidth}
        \centering
        \subfloat[Random negatives]{
            \includegraphics[width=\linewidth]{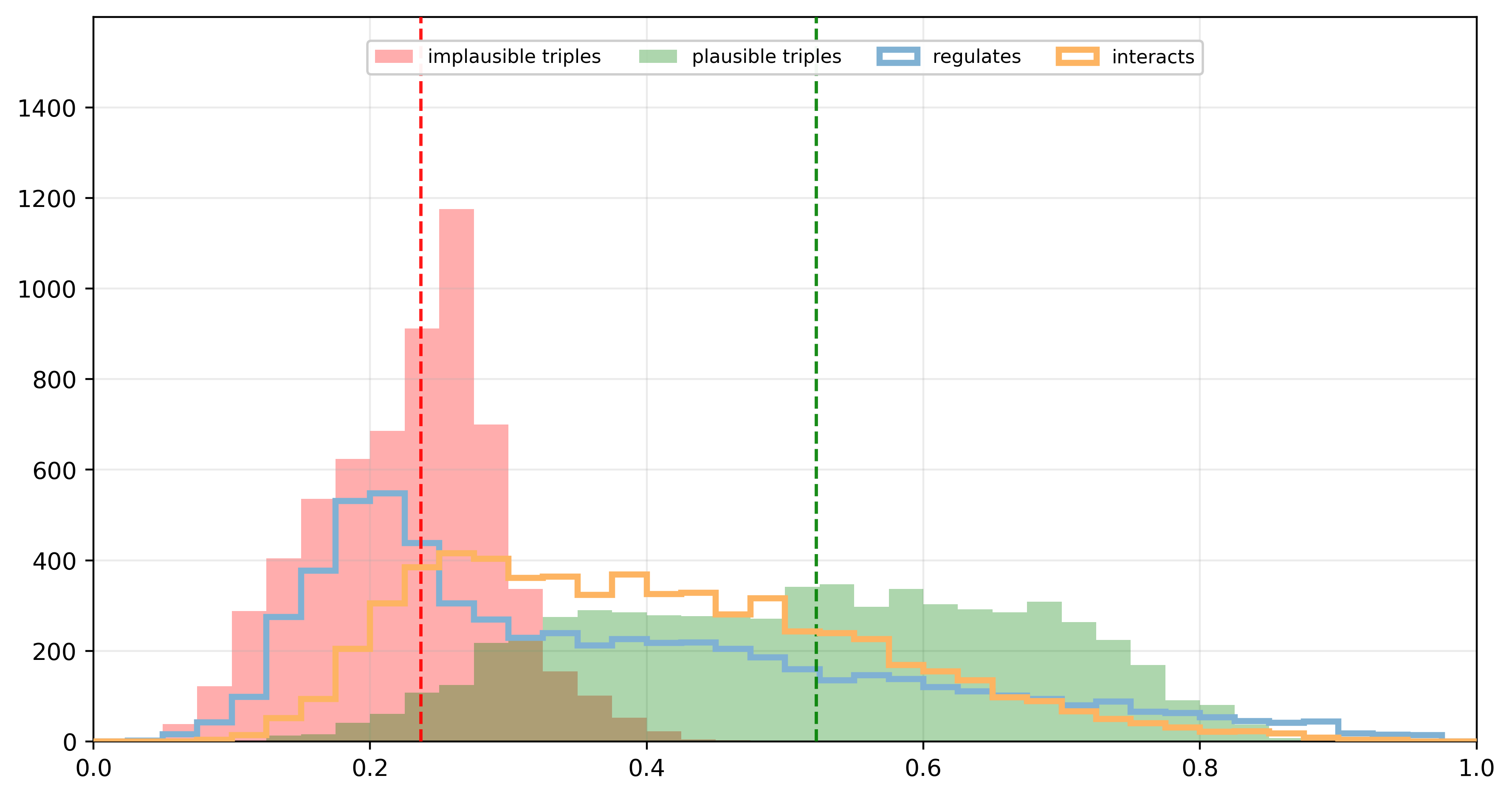}
        }
    \end{minipage}

    \caption{
    Distribution of \(\mathcal{P}_{\tt combo}\) scores for the schema fact \textit{(Gene,covaries,Gene)} in Hetionet.
    }
    \label{fig:app_gene_gene_covaries}
\end{figure}

\begin{figure}[t]
    \centering

    \begin{minipage}{0.46\linewidth}
        \centering
        \subfloat[Community-based negatives]{
            \includegraphics[width=\linewidth]{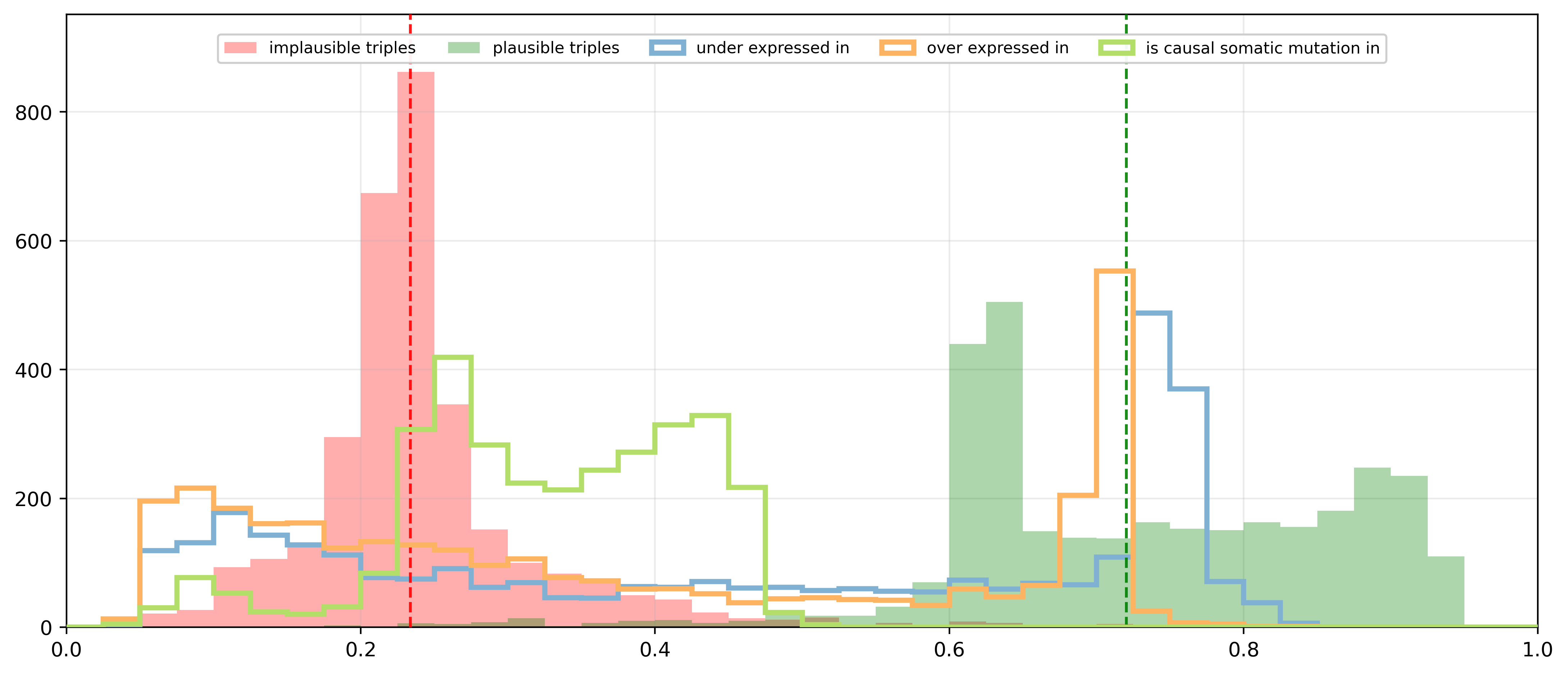}
        }
    \end{minipage}
    \hfill
    \begin{minipage}{0.46\linewidth}
        \centering
        \subfloat[Random negatives]{
            \includegraphics[width=\linewidth]{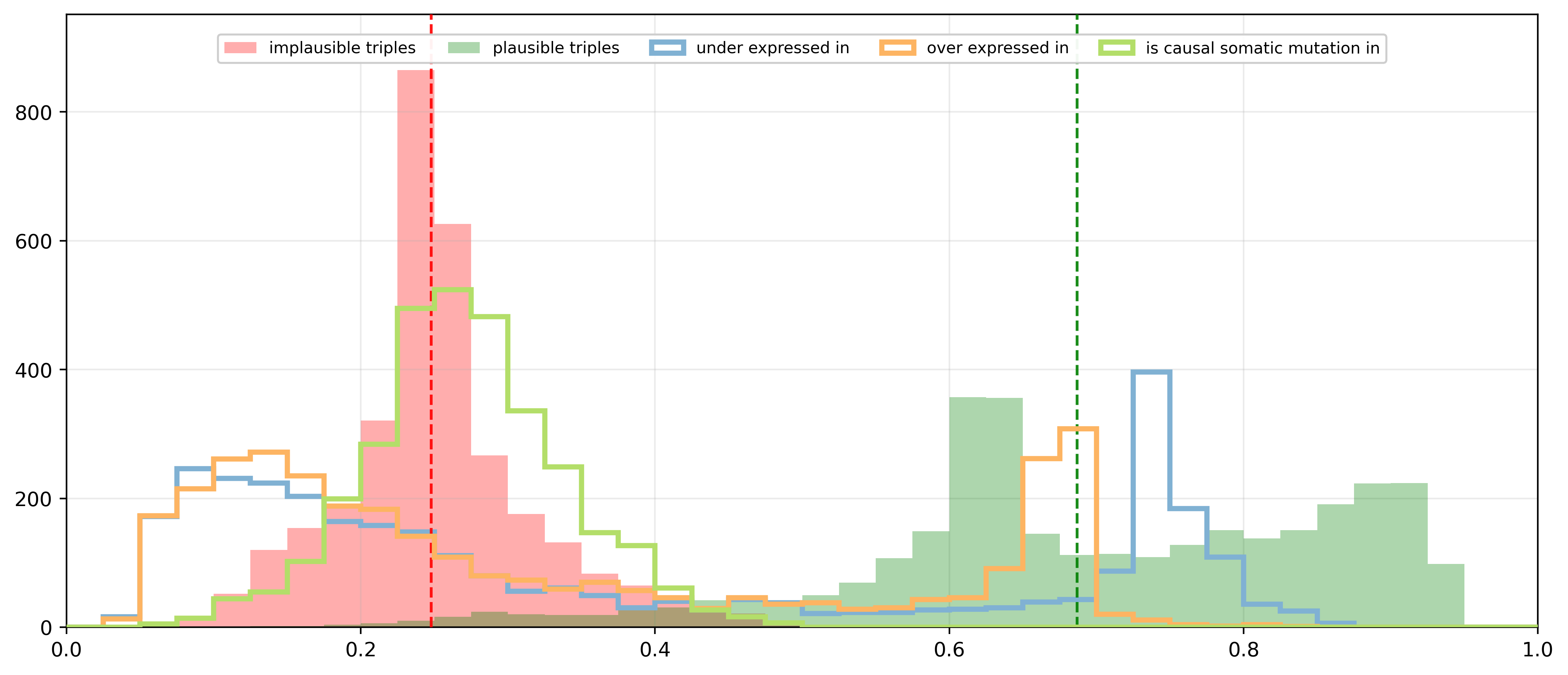}
        }
    \end{minipage}

    \caption{
    Distribution of \(\mathcal{P}_{\tt combo}\) scores for the schema fact \textit{(miRNA,causes or contributes to condition,Disease)} in miRNA-KG.
    }
    \label{fig:app_mirna_causes}
\end{figure}

\begin{figure}[t]
    \centering

    \begin{minipage}{0.46\linewidth}
        \centering
        \subfloat[Community-based negatives]{
            \includegraphics[width=\linewidth]{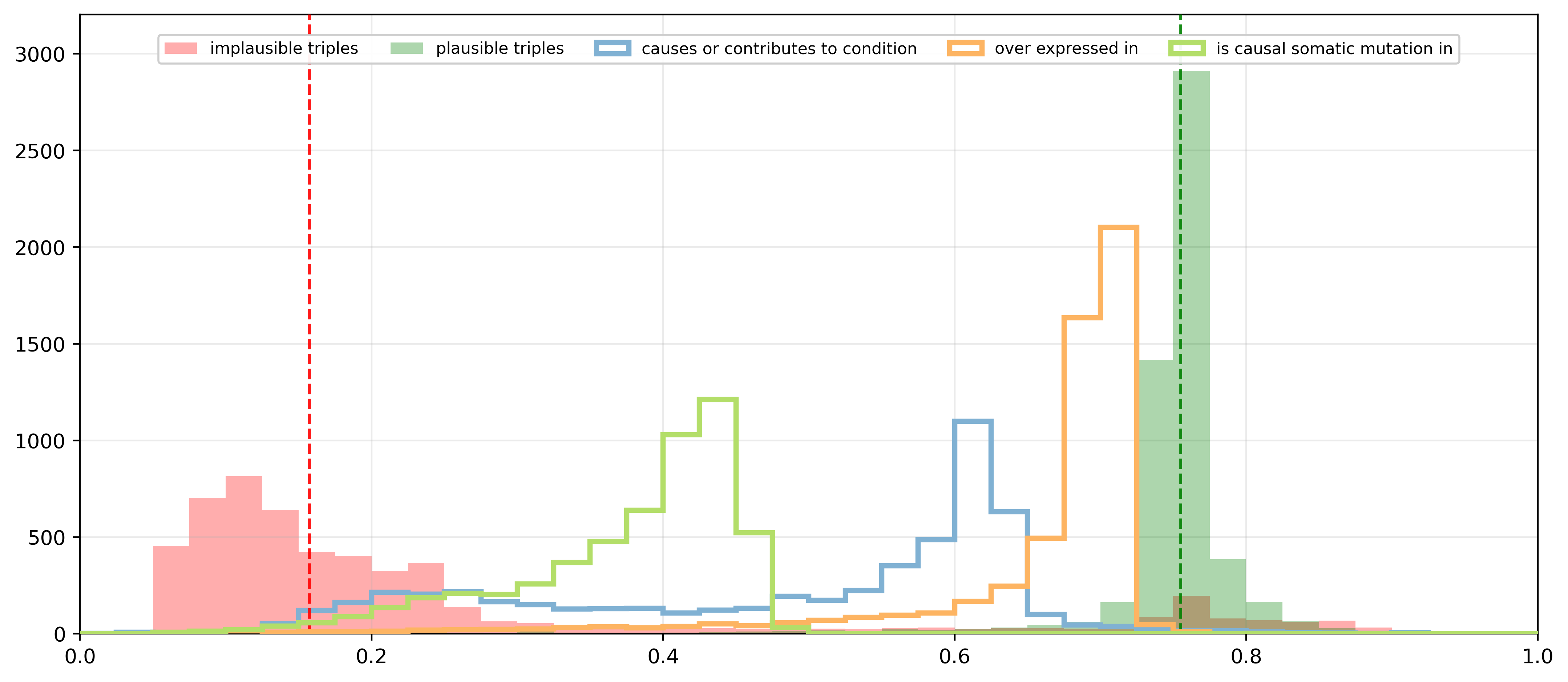}
        }
    \end{minipage}
    \hfill
    \begin{minipage}{0.46\linewidth}
        \centering
        \subfloat[Random negatives]{
            \includegraphics[width=\linewidth]{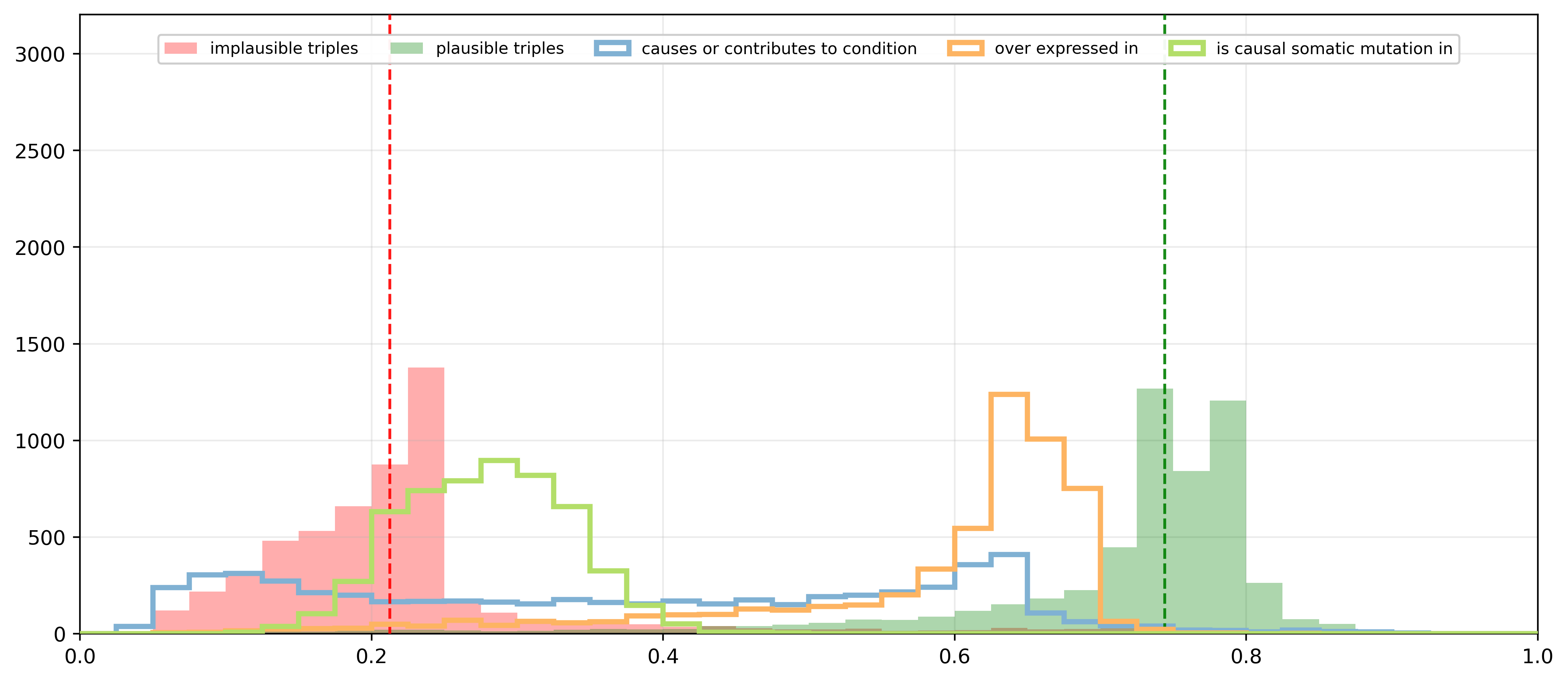}
        }
    \end{minipage}

    \caption{
    Distribution of \(\mathcal{P}_{\tt combo}\) scores for the schema fact \textit{(miRNA,under-expressed in,Disease)} in miRNA-KG.
    }
    \label{fig:app_mirna_underexpressed}
\end{figure}

\section{Additional Qualitative Analysis of Plausibility Distributions}
\label{sec:supp_qualitative}
The quantitative evaluation presented in the main manuscript demonstrates the effectiveness of the proposed plausibility formulations through aggregate calibration and competition metrics. In this section, we provide additional qualitative examples illustrating how \(\mathcal{P}_{\tt combo}\) 
distributions support the interpretation of candidate annotations under different biological contexts.

In all figures, green histograms correspond to blind plausible triples ($BlP_i$), whereas red histograms correspond to generated implausible triples ($N_i$). The blue and orange curves represent the plausibility assigned to competing predicates involving the same entity pair. Vertical dashed lines delimit the three plausibility regions introduced in the main paper, namely \emph{implausible}, \emph{uncertain}, and \emph{plausible}.

\subsection{Disease--Gene interactions in Hetionet}

Figure~\ref{fig:bio_competitors_upregulates} reports the plausibility distributions obtained for the schema fact
\textit{(Disease,upregulates,Gene)}.
Similarly to the example discussed in the main paper, community-based negative
sampling produces substantially clearer score distributions than the random
baseline. The overlap between plausible and implausible triples decreases from
65.1\% to 22.3\%, allowing the proposed plausibility formulation to better
distinguish highly plausible annotations from weakly supported candidates.

An interesting observation concerns the competing predicate \textit{associates}, which consistently receives relatively high plausibility scores. Rather than indicating an incorrect prediction, this behavior reflects the biological relationship between the predicates. Disease--gene associations represent a broader semantic concept that naturally includes more specific regulatory mechanisms such as gene upregulation. Consequently, high plausibility assigned to \textit{associates} should be interpreted as evidence supporting a general biological relationship, while the more specific mechanistic annotation requires further expert validation.

\subsection{Gene--Gene interactions in Hetionet.}
Figure~\ref{fig:app_gene_gene_interacts} reports the distributions for the schema fact \textit{(Gene,interacts,Gene)}. In both strategies, plausible and implausible triples are well separated, but community-based negatives further reduce the overlap between the two distributions, from \(12.6\%\) with random negatives to \(5.4\%\). The competing predicates \textit{regulates} and \textit{covaries} receive intermediate-to-high plausibility scores, which is coherent with the fact that they represent biologically related, but more specific or indirect, gene--gene relations. This behavior confirms that our approach does not simply reject competing predicates, but places them in regions where expert inspection may be useful.

Figure~\ref{fig:app_gene_gene_covaries} shows the complementary case \textit{(Gene,covaries,Gene)}. Community-based negatives lead to a clearer separation between plausible and implausible triples, reducing their overlap from \(18.9\%\) to \(7.2\%\). The competitor \textit{regulates} is mostly shifted toward lower or intermediate plausibility scores, while \textit{interacts} receives broader intermediate-to-high scores. This is consistent with the fact that gene co-variation can be compatible with interaction-like signals, but does not necessarily imply direct regulation.

\subsection{miRNA--Disease interactions in miRNA-KG.}
Figure~\ref{fig:app_mirna_causes} reports the distributions for \textit{(miRNA,causes or contributes to condition,Disease)}. The two strategies both produce a strong separation between plausible and implausible triples, but community-based negatives slightly reduce the overlap between the two distributions, from \(7.7\%\) to \(5.0\%\). Competing predicates  receive distinct score profiles. For example, the predicate \textit{over-expressed in} receives intermediate plausibility scores, suggesting that altered miRNA expression may provide evidence for a disease-related association without necessarily establishing the broader \textit{causes or contributes to condition} relation.

Figure~\ref{fig:app_mirna_underexpressed} shows the case of \textit{(miRNA,under-expressed in,Disease)}. Here, community-based negatives assign a larger fraction of triples to the plausible region (\(29.3\%\) versus \(25.3\%\)) and a smaller fraction to the uncertain region (\(45.7\%\) versus \(49.1\%\)). The competitor curves remain biologically interpretable: \textit{over-expressed in} and related disease--miRNA predicates are not necessarily unrelated alternatives. Indeed, under-expression and over-expression relations may coexist in the bioKG for the same miRNA--disease pair depending on the specific experimental condition, tissue, or biological context considered. For this reason, intermediate scores should not be interpreted as automatic errors, but as cases requiring expert assessment rather than direct rejection.